\documentclass[fleqn,usenatbib]{mnras}

\usepackage{newtxtext,newtxmath}

\usepackage[T1]{fontenc}
\usepackage{ae,aecompl}
\usepackage{xcolor}

\usepackage{graphicx}	
\usepackage{amsmath}	
\usepackage{amssymb}	
\usepackage{subfig}

\newcommand{\ch}{\textcolor{black}}

\title[Changing cloud cover on $\epsilon$ Indi Ba,Bb]{Large-scale changes of the cloud coverage in the $\epsilon$ Indi Ba,Bb system}

\author[J. A. Hitchcock et al.]{
J. A. Hitchcock,$^{1,2}$\thanks{E-mail: jah36@st-andrews.ac.uk}
Ch. Helling,$^{1,2,3}$
A. Scholz,$^{1,2}$
G. Hodosan,$^{1,2}$
M. Dominik,$^{1,2}$
\newauthor
M. Hundertmark,$^{7}$
U. G. J{\o}rgensen,$^{4}$
P. Longa-Pe{\~n}a,$^{17}$
S. Sajadian,$^{9}$ 
J. Skottfelt,$^{21}$
\newauthor
C. Snodgrass,$^{22}$
V. Bozza,$^{5, 6}$
M. J. Burgdorf,$^{10}$
J. Campbell-White,$^{26}$
\newauthor
Roberto Figuera Jaimes,$^{1, 29, 30}$
Y. I. Fujii,$^{27,12,28,4}$
L. K. Haikala,$^{13}$
T. Henning,$^{14}$
\newauthor
T. C. Hinse,$^{15}$
S. Lowry,$^{8}$
L. Mancini,$^{16,14,24,25}$
S. Rahvar,$^{18}$
M. Rabus,$^{19,20}$
\newauthor
J. Southworth,$^{11}$
C. von Essen$^{23}$
(The MiNDSTEp Collaboration)
\\
The authors' affiliations are shown in Appendix \ref{sec:affiliations}
}

\date{Accepted XXX. Received YYY; in original form ZZZ}

\pubyear{2020}

\begin{document}
\label{firstpage}
\pagerange{\pageref{firstpage}--\pageref{lastpage}}
\maketitle

\begin{abstract}
We present the results of 14 nights of \textit{I}-band photometric monitoring of the nearby brown dwarf binary, $\epsilon$ Indi Ba,Bb. Observations were acquired over 2 months, and total close to 42 hours of coverage at a typically high cadence of 1.4 minutes. At a separation of just $0.7''$, we do not resolve the individual components, and so effectively treat the binary as if it were a single object. However, $\epsilon$ Indi Ba (spectral type T1) is the brightest known T-type brown dwarf, and is expected to dominate the photometric signal. We typically find no strong variability associated with the target during each individual night of observing, but see significant changes in mean brightness -  by as much as $0.10$ magnitudes - over the 2 months of the campaign. This strong variation is apparent on a timescale of at least 2 days. We detect no clear periodic signature, which suggests we may be observing the T1 brown dwarf almost pole-on, and the days-long variability in mean brightness is caused by changes in the large-scale structure of the cloud coverage. Dynamic clouds will very likely produce lightning, and complementary high cadence \textit{V}-band and H\textit{$\alpha$} images were acquired to search for the emission signatures associated with stochastic \lq strikes\rq. We report no positive detections for the target in either of these passbands.    
\end{abstract}




\section{Introduction}\label{sec:intro}

As brown dwarfs are not massive enough to maintain stable hydrogen fusion, they will inevitably cool with age. Initially, most brown dwarfs are late M dwarfs, but as they evolve they exhibit spectral types of L, T and Y. As a brown dwarf evolves the molecular gases in its atmosphere will condense and form clouds. These largely consist of \ch{ particles made of a mix of silicate, metallic oxide and iron}, whose refractory properties are encoded into the brown dwarf's spectra (e.g. \citet{burrows2006and}, \citet{helling2008comparison}).

The $\epsilon$ Indi Ba,Bb brown dwarf binary hosts a T1 (Ba) and a T6 (Bb) component \citep{king2010indi}. At a distance of only 3.6 pc, $\epsilon$ Indi Ba is the brightest known T dwarf. Given this potential for precise photometry, this system provides an excellent case study with which to probe the transition from L to T spectral types. During this stage of a brown dwarf's lifetime, it is likely to exhibit pronounced optical variability, associated with changes in cloud coverage above the photosphere (e.g. \citet{radigan2014strong}, \citet{metchev2015weather}, \citet{eriksson2019detection}), with the exact nature of the observable signatures being partly dependent on the inclination angle. For example, rotating spots and weather systems on the surface may be used to infer the rotation periods of these objects when viewed close to the equator. \textcolor{black}{Irregularly evolving time series can result from phase shifts between banded structures on the surface with different periods \citep{apai2017zones}.} Further, the continued aggregation of condensates onto the surface of the cloud particles may cause the clouds to thin \ch{into an optically transparent atmosphere}, and ultimately rain-out into the lower\ch{, optically thick} atmosphere (e.g. \citet{knapp2004near}). \textcolor{black}{Time series photometry has proven to be an invaluable tool for monitoring the atmospheric variability of these objects \citep{apai2019mapping}, and for the past two decades has been successful in measuring both the periodic and quasi-oscillatory signatures seen in both old and young brown dwarfs (e.g. see \citet{scholz2004rotation}, \citet{artigau2009photometric}) with surveys both on the ground \citep{vos2019search} and in space \citep{biller2018simultaneous}.} 

The \textit{I}-band observations presented in this paper were typically acquired at a high cadence of $\sim1.4$ minutes. One motivation for probing the short-timescale variation of this ultracool target is to test for signatures potentially diagonostic of lightning in the atmospheres of the binary's components. Lightning strikes are short, stochastic events which are expected to occur collectively \textcolor{black}{in something like an extrasolar storm \citep{yang2016extrasolar}}. Since lightning has never been observed on extrasolar objects, the typical duration of individual strikes in brown dwarf atmospheres is unknown. However, individual strikes will likely have a sub-second duration, with e.g. an average duration of $10^{-4}$ s on Earth \citep{volland2013atmospheric}, and 0.3 s for the slowest Saturnian strikes \citep{zarka2004study}. This, however, can be very different on brown dwarfs and on exoplanets due to different atmospheric chemistries, dynamics and density structure.

The net behaviour of these strikes can be a brightening of the target \ch{in the optical but also in the radio or UV}, with a magnitude dependent on the brightness temperature of individual strikes, percentage coverage of the storm over the hemisphere, the rate of strikes etc. The effects may also manifest as darkening in specific wavelength bands due to chemical changes caused by lightning (see Table 1 of \citet{bailey2014ionization}), e.g. due to the occurrence of HCN at the expense of CH$_4$ \citep{hodosan2016lightning}, or more complex molecular ions like HCO\textsuperscript{+} suggested in the more rarefied gases of planet-forming disks \citep{helling2016atmospheric}.
The short, stochastic brightening associated with strikes could be probed by quantifying the asymmetry in the flux of the light curve, provided the signature exceeds the noise, of course.

Extrasolar lightning has never been observed, and if lightning is indeed present on $\epsilon$ Indi Ba,Bb, its properties may be very different to that of lightning strikes observed in the Solar System. Planning an ideal strategy to detect strikes is, therefore, challenging. Chapter 7 of \citet{hodosan2017lightning} details a parameter study to estimate a range of possible optical fluxes of lightning strikes originating in the atmosphere of $\epsilon$ Indi Ba,Bb, and subsequently, the feasibility of observing these strikes with the telescope and filter system used in this work (\S \ref{sec:obs+datared}). \textcolor{black}{The parameters considered by \citet{hodosan2017lightning}, and the associated equations, are listed in Appendix \ref{sec:lightningeqs}.} Necessarily, the properties of Solar System lightning must be used, but it is shown that if lightning strikes in the brown dwarf's atmosphere occur over its hemisphere with a flash density (i.e. the rate of strikes per unit area) comparable to that which is observed in the plumes of volcanic eruptions on Earth, then these strikes would cause an increase in brightness similar to that of the combined brightness of both brown dwarfs, and be easily observed in this study. \textcolor{black}{To take the most promising case as just one example, if we assume the strikes have power and discharge durations as estimated by \citet{bailey2014ionization}, i.e. (\S\; Equations \ref{eq:Iobs} - \ref{eq:pogson}) $P_{\textrm{opt,fl}}=10^{15}$ W, $\tau_{\textrm{fl}}=10^{-4}$ s, and occur with a flash density like that observed during the Mt Redoubt Eruption (2009-03-29), $\rho_{\textrm{fl}}$=2000 km\textsuperscript{-2}.h\textsuperscript{-1}, over the brown dwarf's entire visible hemisphere, this would result in a signal with an apparent magnitude of about 15.2 and 15.7 in the \textit{I} and \textit{V} bands respectively. It is thought that the volcanic dusty plumes associated with these very high flash densities may be analogous to the silicate rich dust clouds present in brown dwarf atmospheres \citep{helling2008comparison}.} \textcolor{black}{A full exploration of the parameter space for the properties of extrasolar lightning on $\epsilon$ Indi Ba,Bb is outside the scope of this work, and so we refer the reader to Table 7.6 in Chapter 7 of \citet{hodosan2017lightning} for a summary of possible signal strengths.}

At a separation of just $0.7''$, the individual components of this binary are rarely resolved with conventional ground-based imaging. As such, in the work presented here, we measure the photometry for the combined system. This is true for all previous studies of the time-varying brightness of this source, which we discuss below.

Following the discovery of the system \citep{scholz2003varepsilon}, \citet{koen2003search} obtained 2.3 and 3.3 hours of \textit{I} band photometry on two nights, 4 days apart. A drop in mean magnitude of $\sim 0.1$ mag is seen between the two nights, in addition to an enhanced scatter relative to the comparison stars in each night's time series. A gradual brightening of the target is also seen in both nights, and a $0.05$ magnitude (mag) linear rise is seen over about 3 hours.

Soon afterwards, the binary nature of the system was established \citep{mccaughrean2004varepsilon}. \citet{koen2004search} revisited the target, acquiring 2.9 hours of \textit{H}- and \textit{K}-band photometry. Aperiodic scatter, albeit no greater than at the level of 12 mmag is seen in both bands. A period of 3.1 hr is shown to fit the concurrent \textit{H} and \textit{K} photometry well, yet given this exceeds the length of the run, the authors point out that this is clearly not a reliable detection.

\citet{koen2005ic} presents 3.6 hours of \textit{I\textsubscript{C}} photometry of the target, over which a strong linear brightening is observed, at a rate of $0.75$ magnitudes per day (i.e. ~0.1 mag over the course of the run). A plot of the residuals of a straight line fit to the trend suggests an additional random component to the variability, which with reference to comparison stars, does not likely arise from atmospheric or instrumental effects. This suggested that the 3.1 hour period was either transient or incorrect, and supported the variability first described in \citet{koen2003search}. Indeed, these results taken together suggested that $\epsilon$ Indi Ba,Bb shows comparable levels of variability on both days-long and hourly timescales. With follow-up \textit{K\textsubscript{S}} differential photometry, \citet{koen2005jhk} further showed the target to have faded by $\sim0.05-0.06$ mag between June 2003 and October 2004. 


Point spread function (PSF) photometry was the chosen approach for all of these photometric studies. As discussed in \citet{koen2009correcting}, for unresolved binary systems, stellar magnitudes determined by PSF fitting may be systematically affected by seeing. Therein, it is suggested this effect arises from seeing-dependent PSF variations when observing unresolved binaries with angular separations just below the best resolution limit. With a component separation of roughly $0.7''$, the prior PSF fitting photometry of $\epsilon$ Indi Ba,Bb is expected to be strongly affected by this systematic effect. Indeed, \citet{koen2012extensive} revisits the observations of the target -- in addition to obtaining three new sets of $\sim3.0$ hour \textit{I}-band observations -- and finds that all the time series show a strong dependence of variability with seeing. Correcting for this, the linear rise seen in \citet{koen2005ic} is replaced with a very different, smoothly varying function, and over the now four runs that do in fact show short time-scale brightness changes, the revised level of variability is shown to be much smaller than previously suggested, with only two nights showing a variation as large as 0.05 mag over a few hours. The conclusion of significant differences in mean brightness between nights however still holds, with a range of 0.136 mag over this decade of intermittent \textit{I}-band observations.

In this paper, we present the results of 14 nights of high-cadence, \textit{I}-band photometric monitoring of the combined $\epsilon$ Indi Ba,Bb system, acquired over 2 months. In Section \ref{sec:obs+datared} we outline the observations, data reduction and the approach for the differential photometry. The subsequent results are presented in Section \ref{sec:genvariability}, followed by a discussion of their astrophysical implications in Section \ref{sec:discussion}. We summarise our conclusions in Section \ref{sec:conclusion}.

\begin{table*}
	\centering
	\caption{Spectral classification, effective temperature, bolometric luminosity and apparent \textit{I}- and \textit{V}-band magnitudes for each component of the $\epsilon$ Indi Ba,Bb system \citep{king2010indi}. (a) Temperature derived by fitting atmospheric models to the observed spectra.}
	\label{tab:example_table}
	\begin{tabular}{cccccc}
		\hline
		Component & Spectral Type & $T_{\mathrm{eff}}$\textsuperscript{a} [K] & lg $L/L_\odot$ & \textit{I}-band [magnitude] & \textit{V}-band [magnitude]\\
		\hline
		$\epsilon$ Indi Ba & T1 & 1300-1340 & $−4.699 \pm
0.017$ & $17.15 \pm 0.02$ & $24.12 \pm 0.03$\\
		$\epsilon$ Indi Bb & T6 & 880-940 & $ −5.232 \pm
0.020$ & $18.921 \pm 0.02$ & $\geq26.60 \pm 0.05$\\
		\hline
	\end{tabular}
	\label{tab:epsiinfo}
\end{table*}

\section{Observations and Data Reduction}\label{sec:obs+datared}
\subsection{Data acquisition and pre-processing}

\textit{I}-band observations of $\epsilon$ Indi Ba,Bb were acquired on 15 separate nights between 2017-07-24 and 2017-09-24 with the Danish 1.54m telescope (DK1.54) at ESO La Silla, as part of the 2017 MiNDSTEp\footnote{http://www.mindstep-science.org/} season. A total of 1457 good\footnote{Good images are those which have not been flagged for spurious photometry (see Section \ref{sec:comparstar})} \textit{I}-band images were acquired with the Danish Faint Object Spectrograph Camera (DFOSC), which uses a 2K$\times$2K thinned Loral CCD chip with a pixel scale of $0.4''$ per pixel, giving a FoV of $~13.7' \times 13.7'$. Typically, an exposure time of 60 s was used, providing a rapid observational cadence of about 1.4 minutes. The observing log -- which includes the exposure times and total length of each nightly run -- is shown in Table \ref{tab:obsinventory}. Unfortunately, due to high winds, observations were terminated prior to finding the correct pointing on the night of 2017-08-03, and so we exclude these 5 frames from the analysis (see Section \ref{sec:comparstar}).

The images were reduced with a bias subtraction and flat-field correction. For most nights, $\sim10$ biases and $~10$ \textit{I}-band flats were obtained. All calibration frames were averaged (median), and the master bias and flats were used to reduce the data. For a small number of nights when calibration frames were not obtained, there were always suitable flats and biases acquired either the previous or following day that could be used for the reduction. The {\sc Source Extractor} \citep{bertin1996sextractor} software was used to perform the aperture photometry.

\begin{table}
	\centering
	\caption{Observation inventory for the entire observing campaign. (a) Exposure time.}
	\label{tab:example_table}
	\begin{tabular}{p{2cm} p{1.5cm} p{1cm} p{1.7cm}} 
       \hline
       Date (YYYY-MM-DD) & Number of 'good' images & $t_{\mathrm{exp}}$\textsuperscript{a} [s] & Duration of run [hr]\\
       \hline\\
       2017-07-24  &  38 & 120 & 1.54 \\
       2017-08-03  &  5 & 60 & 0.15\footnotemark\\
       2017-08-15  &  81 & 60 & 3.58\footnotemark \\
       2017-08-17  &  56 & 60 & 1.97 \\
       2017-08-20  &  70 & 60 & 2.02\footnotemark \\
       2017-08-28  &  165 & 60 & 4.07 \\
       2017-08-30  &  43 & 80 &  2.75 \\
       2017-09-03  &  60 & 60 & 1.40 \\
       2017-09-05  &  128 & 60 & 3.74 \\
       2017-09-13  &  138 & 60 & 3.98 \\
       2017-09-15  &  166 & 60 & 3.97 \\
       2017-09-17  &  165 & 60 & 3.99 \\
       2017-09-20  &  165 & 60 & 4.04 \\
       2017-09-22  &  118 & 60 & 3.23 \\
       2017-09-24  &  143 & 60 & 3.50 \\
		\hline
	\end{tabular}
	\label{tab:obsinventory}
\end{table}

\footnotetext[3]{Strong winds meant observations were prematurely aborted. We exclude this night from any further analysis.}
\footnotetext[4]{\textcolor{black}{Both cloudy conditions and suspected light contamination on the chip from the nearby, bright companion star, $\epsilon$ Indi A, produced spurious photometry on this night, resulting in $\sim1$ hr of coverage being clipped. This latter issue is unique to the photometry on this night only (\S \ref{sec:comparstar}).}}
\footnotetext[5]{Partial interruption of $\sim0.5$ hr, probably due to cloud.}

To complement the long-term \textit{I}-band monitoring, intermittent \textit{V}-band and H\textit{$\alpha$} observations were acquired at a similarly high cadence with a consistent, 60 s exposure time. Under normal conditions, we do not expect the target to be observable above the background in these passbands with this relatively short exposure time. Rather, the motivation for these observations was to search for signatures that may be associated with stochastic lightning strikes separate from the continuum spectrum of the ultracool target (i.e. the necessarily strong atomic and molecular emission lines at these shorter wavelengths). In Section \ref{sec:nolightning}, we discuss the results of an analysis of 234 \textit{V}-band and 412 H\textit{$\alpha$} images.

\subsection{Comparison star selection}\label{sec:comparstar}

An ensemble of stable comparison stars was found with application of the following criteria to each night of observations: 1) The star must be within 2 magnitudes of the target\footnote{Brighter stars risk producing diffraction spikes on the CCD}; 2) The standard deviation of the star's magnitudes must be less than that of the target and 3) The star must appear on at least as many images as the target. This provided a candidate set of 16 comparison stars.

The differential photometry for each night was performed by normalising the raw comparison star time series by their weighted mean magnitude, and taking the mean across all of the subsequent residual time series. The residual magnitudes may then be subtracted (i.e. a division in flux) from the raw time series of all stars in the FoV. The corrected time series of the comparison stars were then inspected by eye for any signs of variability, and no further cuts were deemed necessary.

To guard against the impact of spurious images on the differential photometry, a $5\sigma$ clip was applied to the raw time series of the comparison stars. All sigma clipping described in this work -- applied to both the raw and differential photometric time series -- is done with respect to the median absolute deviation (MAD) of the scatter in the time series, scaled to a standard deviation,

\begin{equation}
    \sigma_{\text{MAD}} = 1.4826 \times \text{median}(|m_i - \Tilde{m}|)\;,
    \label{eq:sigmad}
\end{equation}
for magnitudes $m_i$ with median $\Tilde{m}$. The factor of 1.4826 is the scaling required for normally distributed data.

It is required that the same ensemble of comparison stars is used for the differential photometry of every data point in our \textit{I}-band time series. This means that if the photometry of just a single member of the ensemble is flagged by the {\sc Source Extractor} software for any given frame, the differential photometry for that frame will not be calculated. \textcolor{black}{Typically, no more than 3 per cent of all frames acquired on a given night were rejected, with the notable exceptions of the nights of 2017-08-15 and 2017-08-30 -- where the quality of the photometry was severely affected by intermittent cloud coverage for the former, and both cloud coverage and suspected light contamination by $\epsilon$ Indi A for the latter -- and the night of 2017-08-17, where stars were lost due to an initially inaccurate pointing.}

\begin{figure}
    \centering
    \includegraphics[width=\columnwidth]{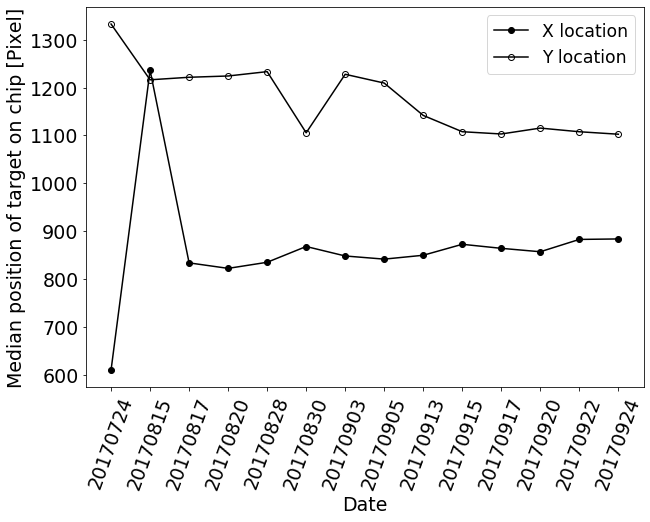}
    \caption{The median X and Y pixel locations for $\epsilon$ Indi Ba,Bb for each night of \textit{I}-band photometry analysed in this work. Following the night of 2017-08-15, a consistent, more easterly pointing was used to ensure the bright companion star, $\epsilon$ Indi A, was not falling directly on the chip.}
    \label{fig:targetlocation}
\end{figure}

\textcolor{black}{The bright nearby companion star, $\epsilon$ Indi A, was visible on the chip for most of the night of 2017-08-15. $\epsilon$ Indi A is located west of the target, along the X-axis of the CCD, and for the remainder of the campaign care was taken to ensure this object was not directly falling on the chip. This is highlighted by Figure \ref{fig:targetlocation}, in which we plot the median pixel locations for the target, $\epsilon$ Indi Ba,Bb,  along the X and Y axes of the chip (as a proxy for the pointing), for each night of \textit{I}-band photometry analysed in this work.}

{}

Over all nights of the campaign, we measure a median $\sigma_{\text{MAD}}$ of 23 mmag for the target (see Figure \ref{fig:mints_vs_simgmad}).

\subsection{Calculation of zero-point offsets}\label{sec:zpcalc}

In addition to allowing an examination of variability on minute- to hour-long timescales, the time coverage of this data set allows an investigation into the variability of $\epsilon$ Indi Ba,Bb over almost 2 months. In order to do this, one must calculate the photometric zero-point between different nights of observing to account for the systematic changes in conditions. The simple, heuristic approach used here, is to calculate the mean shift in magnitude of the reference stars relative to the first night, and apply this shift (i.e. an approximation to the zero-point) to all stars. Here, the difference in mean magnitudes between the first night and night $j$ for comparison star $k$ is written as $d_{jk}$. The mean shift in brightness over $K$ comparison stars, and the corresponding sample variance, is then equal to

\begin{equation}
    \overline{d_j} = \frac{1}{K}\sum_{k=1}^{K}d_{jk},\;\; {S_j}^2 = \frac{1}{K-1}\sum_{k=1}^{K}(d_{jk}- \overline{d_j})^2.
    \label{eq:zp}
\end{equation}

It is not guaranteed that reference stars stable over hour-long timescales are stable over many days, and indeed, two stars were discarded from the 16-large ensemble for calculating zero points. $\epsilon$ Indi Ba,Bb is a very red source, and so second-order colour effects - which are associated with the wavelength-dependent nature of atmospheric extinction - may be an issue for this target. Encouragingly however, no clear trend with colour was seen in the reference stars when calculating the zero-point shifts, nor was enhanced scatter in red sources (including the target) seen within nights\footnote{Colour and coordinate information for sources in the FoV was provided by Gaia DR2 \citep{2016}}.

\section{General Assessment of Variability}\label{sec:genvariability}

\subsection{Within-night variation}\label{sec:withinnight}

\begin{figure}
	\includegraphics[width=\columnwidth]{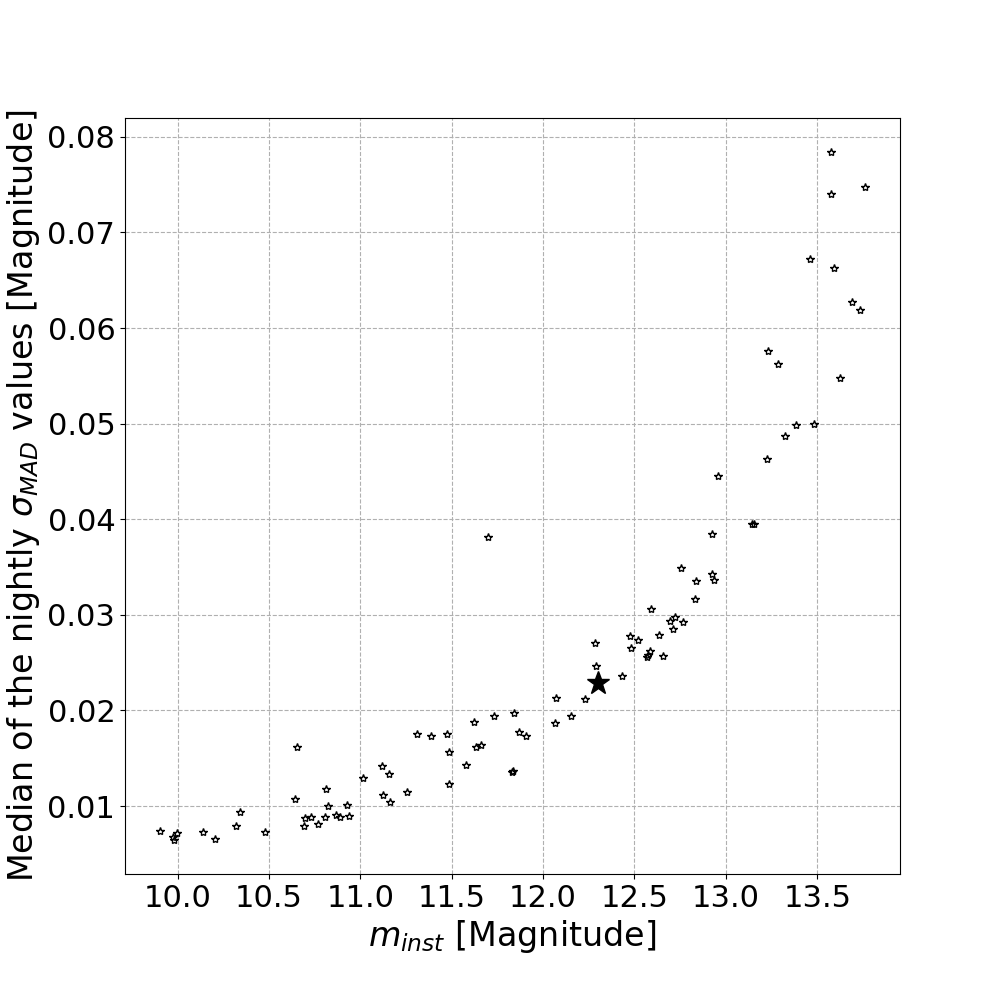}
    \caption{Median instrumental magnitude for $\epsilon$ Indi Ba,Bb (large filled star) and other stars in the FoV (small hollow stars) plotted against their median nightly scatter over 14 nights of observations. A variable star with a $m_{\text{inst}} \sim 11.7$ is suggested by the plot, and inspection of the within-night time series of this source showed clear, roughly linear rises and falls in brightness of about 0.1 mag over the longest, $\sim 4$ hour runs.}
    \label{fig:mints_vs_simgmad}
\end{figure}{}

We plot the median instrumental magnitude for the entire zero-point corrected time series against the median nightly scatter in Figure \ref{fig:mints_vs_simgmad}. The zero-point of the DK1.54 is not well defined, and so the instrumental \textit{I}-band magnitudes shown here are placed on an arbitrary scale. The $\sigma_{\text{MAD}}$ statistic is the median absolute deviation of the photometry for each night, scaled to a standard deviation (described by Equation \ref{eq:sigmad}). $\epsilon$ Indi Ba,Bb is shown with the large, filled black star, and all other sources in the field (both comparison stars and field stars) are shown as small hollow stars. This does not suggest any typically enhanced variability on the hours-long timescales of the individual runs.

\begin{figure}
\centering
\subfloat{
\label{fig:first}
\includegraphics[width=0.95\columnwidth,height=2.0in]{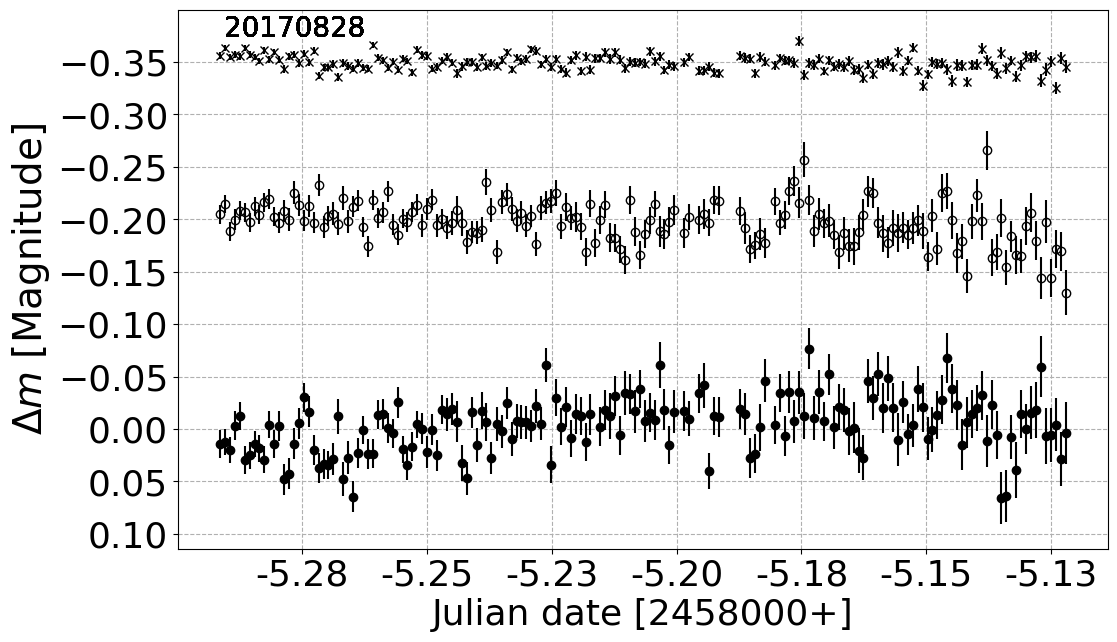}}
\qquad
\subfloat{
\label{fig:second}
\includegraphics[width=0.95\columnwidth,height=2.0in]{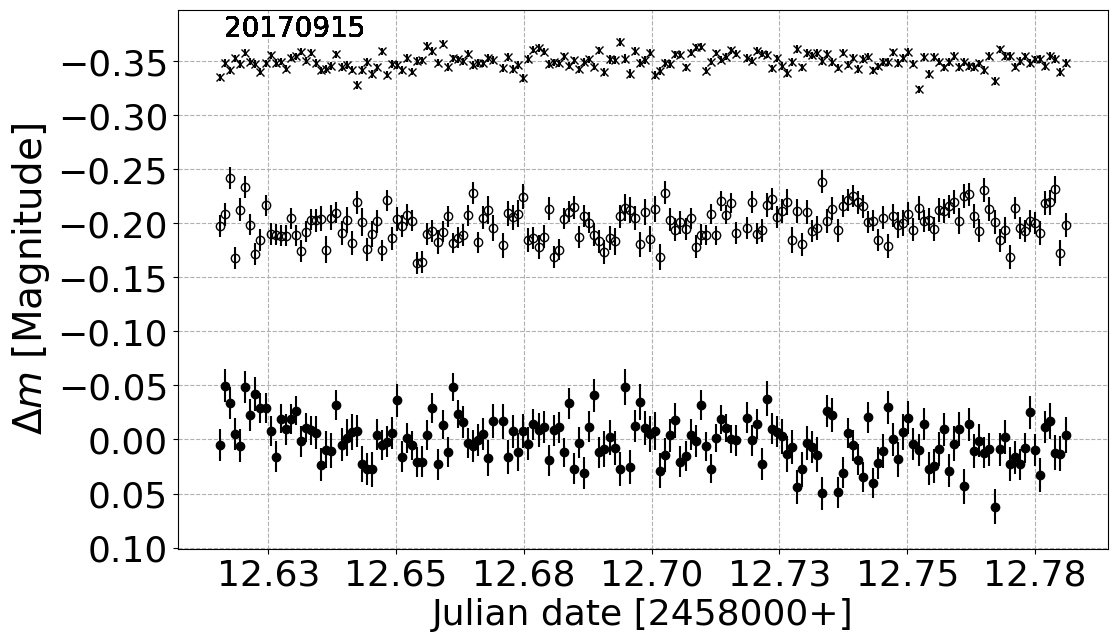}}
\qquad
\subfloat{
\label{fig:third}
\includegraphics[width=0.95\columnwidth,height=2.0in]{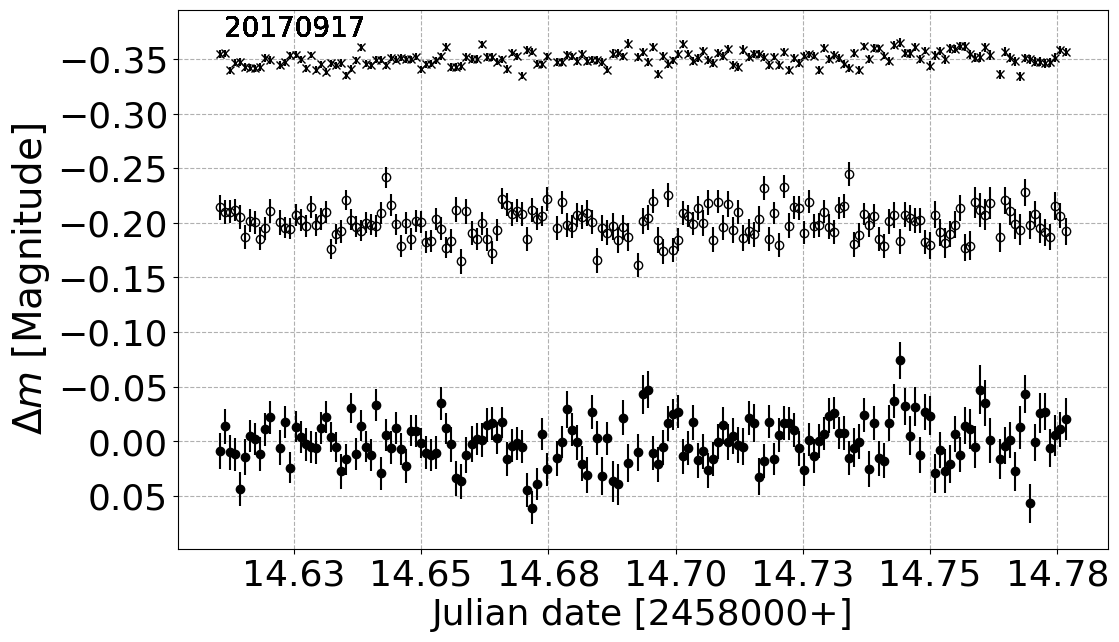}}
\qquad
\subfloat{
\label{fig:fourth}
\includegraphics[width=0.95\columnwidth,height=2.0in]{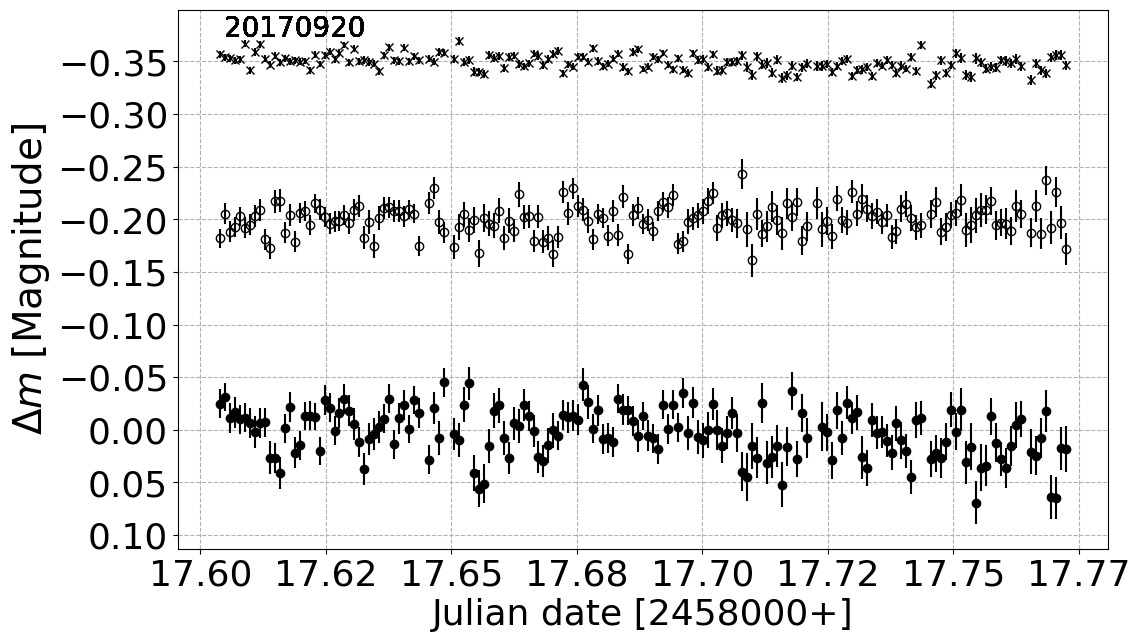}}
\caption{Normalised, arbritrarily offset differential photometry for $\epsilon$ Indi Ba,Bb (filled circles), a red comparison star (hollow circles) and a bright comparison star (crosses) on four different nights. The corresponding dates are shown in the top left of each panel as YYYYMMDD.}
\label{fig:epsilon_timeseries}
\end{figure}

The normalised differential photometry for the four nights with the best data coverage are shown in Figure \ref{fig:epsilon_timeseries}. The target is at the bottom of each of the four panels, represented by the filled circles. For comparison, we plot the light curves for a bright comparison star on the top row (crosses) $~2$ magnitudes brighter than the target, and a red comparison star of comparable brightness to the target (open circles); see Table \ref{tab:starsinfo} for colour and magnitude information. By-eye inspection suggests no greatly enhanced variation in the target relative to the comparison star of similar brightness.

\begin{table*}
	\centering
	\caption{(a) Median instrumental magnitude and associated median absolute deviation (scaled to a standard deviation), for the entire zero-point corrected time series, (b) Gaia DR2 colour.} 
	\label{tab:example_table}
	\begin{tabular}{cccc}
		\hline
		Source & Key & $m_{\text{inst}}$\textsuperscript{a} & $(BP-RP)$\textsuperscript{b}\\
		\hline
		Bright comparison star & Crosses & $10.144 \pm 0.008$ & 1.21\\
		Red comparison star & Open circles & $11.730 \pm 0.017 $ & 2.79\\
		$\epsilon$ Indi Ba,Bb & Filled circles & $12.300 \pm 0.037$ & 6.16\\
		\hline
	\end{tabular}
	\label{tab:starsinfo}
\end{table*}

One can however see short-timescale correlated features in the target time series (e.g. see the roughly 15 minute long \lq peaked\rq\; features on the night of 2017-09-17). However, such features are characteristic of correlated noise, which is expected to affect this particularly red source.

\begin{figure}
    \centering
    \includegraphics[width=0.95\columnwidth]{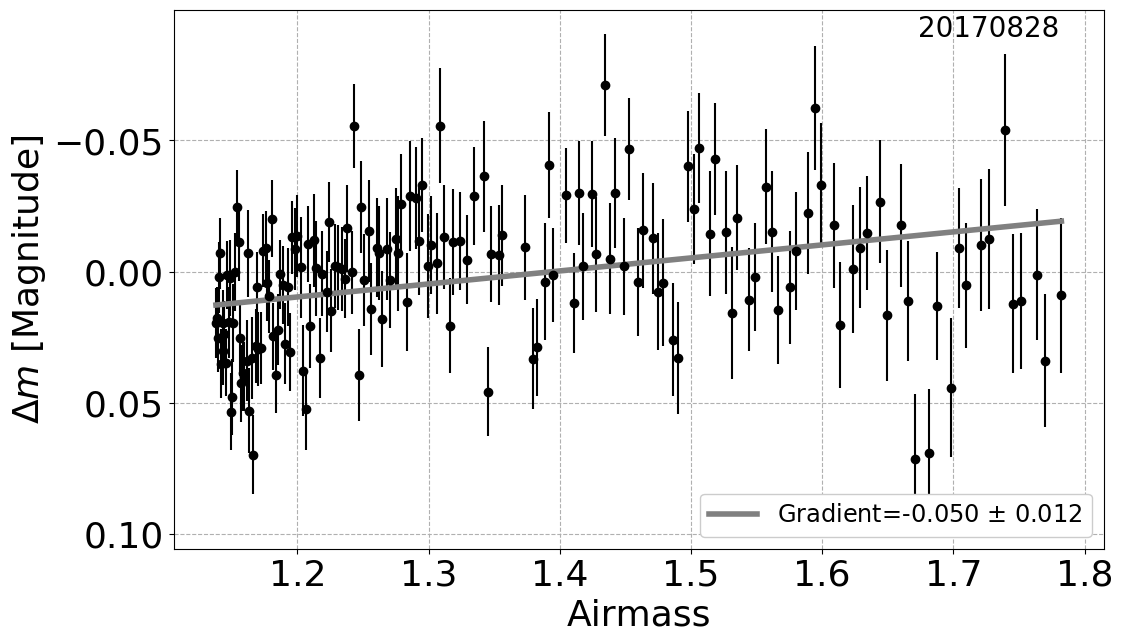}
    \caption{Airmass vs. the normalised differential photometry of the target on the night of 2017-08-28. The gradient and corresponding uncertainty of the plotted straight line fit to the data, and the Pearson's Correlation coefficient, are shown in the legend.}
    \label{fig:Airmass_20170828}
\end{figure}

\begin{figure}
    \centering
    \includegraphics[width=0.95\columnwidth]{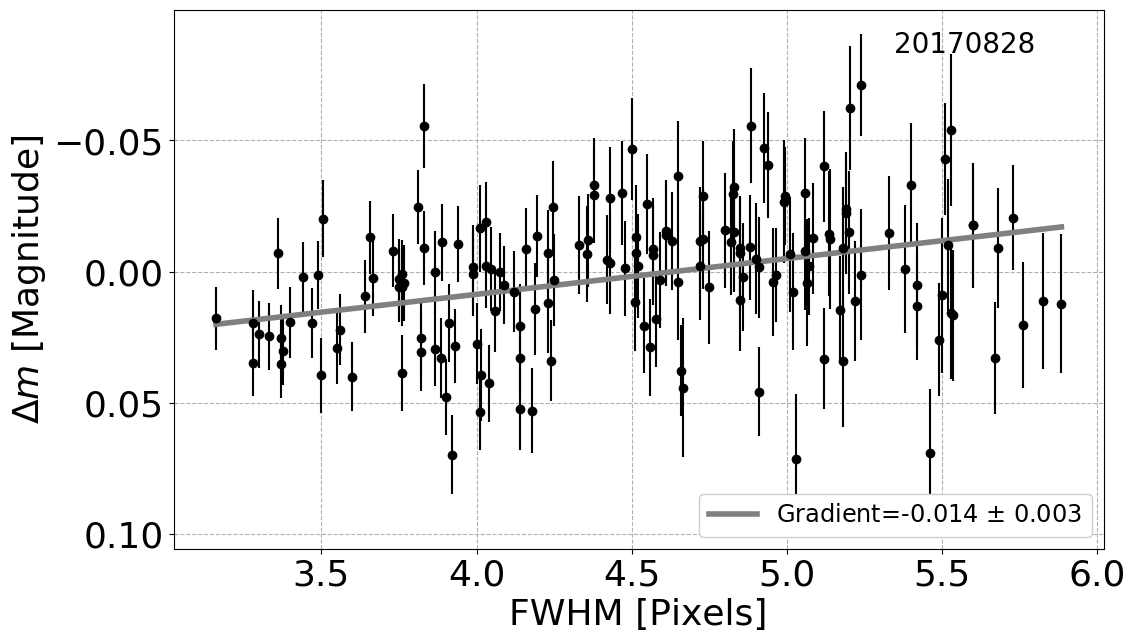}
    \caption{Image FWHM (as a proxy for seeing) vs. the normalised differential photometry of the target on the night of 2017-08-28. The gradient and corresponding uncertainty of the plotted straight line fit to the data, and the Pearson's Correlation coefficient, are shown in the legend.}
    \label{fig:Seeing_20170828}
\end{figure}

\textcolor{black}{The strongest suggestion of any true within-night variability of this target was seen on the night of 2017-08-28 (top panel in Figure \ref{fig:epsilon_timeseries}), where a gradual rise of about 0.05 mag over the first 2 hr is apparent, over which the comparison star time series are flat. In order to ascertain whether this trend could be due to systematic effects, we plot the differential photometry against airmass and image FWHM (as a proxy for the seeing) in Figures \ref{fig:Airmass_20170828} and \ref{fig:Seeing_20170828} respectively. We fit a straight line to each plot with the usual direct least-squares approach. It is not however immediately clear that the relationship between these variables should be linear, and so we empirically estimate an uncertainty for the fit gradient, $\sigma_m$ with $M=1000$ bootstrap trials, $j$, such that}

\begin{equation}
    \sigma_{m}^2 = \frac{1}{M}\sum_{j=1}^{M}[m_j - m]^2\;,
    \label{eq:bootstrap}
\end{equation}

\textcolor{black}{where $m$ is the best-fit gradient when using all the data. This best fit parameter and corresponding uncertainty are shown on the graph. Both Figures suggest the photometry is weakly correlated with both airmass and seeing, and the slopes in both instances are signficantly different from zero.}

\textcolor{black}{To check whether this correlation may be causal, we calculate both correlation coefficients and best-fit slopes for the two comparison stars listed in Table \ref{tab:starsinfo} and directly compare these against the corresponding values for the target. Specifically, we calculate values for both the Pearson correlation coefficient (PCC), and the Spearman rank correlation coefficient (SCC). In order to estimate the uncertainty on the values of these correlation coefficients, we use a Monte Carlo bootstrapping approach to estimate their probability distributions \citep{curran2014monte}. To account for the measurement uncertainties in the photometry (which we assume to be normally distributed), for each of the $M=1000$ bootstrap trials, we perturb the magnitude measurements by adding the measurement uncertainty, $\Delta m_i$, multiplied by a number, $\mathcal{G}$, randomly drawn from a Gaussian of mean 0 and variance 1. For each trial, this is done independently for each magnitude measurement, $m_i$, such that $m_{i, \textrm{perturbed}} = m_i + \mathcal{G}\times\Delta m_i$. As correlation coefficients are bounded between [-1, 1], their sampling distributions will in general be skewed, and so in this work, we state the median value from the empirically estimated probability distributions as our best estimate of the correlation coefficient, with uncertainties corresponding to the upper and lower 34 percentiles of these distributions.}

\textcolor{black}{We note here that the value of the PCC should, in general, be taken with caution for two crucial reasons 1) Its calculation assumes the relationship between the variables under investigation really is linear, and 2) It is very sensitive to outliers. Unlike the PCC, the SCC has the advantage of being a measure of \textit{any} monotonic relationship between the two variables, and is far less sensitive to outliers, and so is perhaps the more accurate diagnostic for assessing the strength of any correlation in a situation where second-order colour effects might be present.}


\textcolor{black}{In contrast to the target, the comparison stars show weak \textit{positive} trends with airmass and seeing. The correlation coefficients and gradients for each star are shown in Tables \ref{tab:AirmassCCtable20170828} and \ref{tab:FWHMCCtable20170820}. The similar correlation coefficients for the two comparison stars suggest that systematics may be influencing our time series for this night, but in the \textit{opposite} way to how the target appears to be varying. That is, it is possible that we may be in fact underestimating the extent of the brightening over the first part of the night.}

\textcolor{black}{On most night however, the target time series was much more weakly correlated with airmass and seeing, with the typical uncertainty on the gradient being of the same order of magnitude as the gradient itself. Nonetheless, significant, albeit very shallow non-zero gradients are apparent. This is not unexpected, since the comparison stars are necessarily bluer than the target. Importantly though, there is no \textit{consistent} positive or negative linear trend with either airmass or seeing on each night. The plots for the target's differential photometry vs Airmass and FWHM for these remaining nights are in Appendix \ref{sec:suppcorrplots}.}

\textcolor{black}{We tabulate these best-fit slopes alongside the SCC for both the target and a comparison star in Tables \ref{tab:AirmassCCtable} and \ref{tab:FWHMCCtable}\footnote{For the reasons discussed above, we drop the PCC from these tables i.e. the relationship between the differential photometry and the parameters under investigation is not necessarily linear.}. That the SCCs for the comparison star on the majority of nights are consistent with, or very close to 0 within the upper and lower 34 percentiles about the medians of the empirically estimated distributions, suggests our approach to differential photometry effectively removes these systematic effects. Typically, the target photometry is also weakly correlated with either seeing or airmass. There are however a few nights where significant SCCs for the target are apparent, but not for the comparison star. As discussed above, these may too be associated with second-order colour effects, but we cannot rule-out the possibility that the target shows small levels of intrinsic variability on these nights e.g. the nights of 2017-09-20 and 2-17-09-24}


{ 
\renewcommand{\arraystretch}{1.5}
\begin{table*}
    \centering
    \begin{tabular}{c|c|c|c}
    \hline
    Star & Slope [vs Airmass] & PCC [vs Airmass] & SCC [vs Airmass]\\
    \hline
    Bright comparison star&$0.012\pm0.003$& $0.25^{+0.09}_{-0.09}$&$0.24^{+0.08}_{-0.08}$\\    
    Red comparison star&$0.037\pm0.008$& $0.30^{+0.10}_{-0.09}$&$0.29^{+0.09}_{-0.09}$\\
    $\epsilon$ Indi Ba,Bb&$-0.050\pm0.012$& $-0.18^{+0.10}_{-0.10}$&$-0.27^{+0.08}_{-0.09}$\\
    \hline
    \end{tabular}
    \caption{Table showing the Pearson (PCC) and Spearman (SCC) correlation coefficients and best-fit slope for the photometry against airmass on the night of 2017-08-28 for the target, and two comparison stars.}
    \label{tab:AirmassCCtable20170828}
\end{table*}
}

{ 
\renewcommand{\arraystretch}{1.5}
\begin{table*}
    \centering
    \begin{tabular}{c|c|c|c}
    \hline
    Star & Slope [vs FWHM] & PCC [vs FWHM]& SCC [vs FWHM]\\
    \hline
    Bright comparison star&$0.004\pm0.001$& $0.31^{+0.07}_{-0.09}$&$0.29^{+0.09}_{-0.08}$\\    
    Red comparison star&$0.009\pm0.002$& $0.27^{+0.09}_{-0.09}$&$0.28^{+0.09}_{-0.09}$\\
    $\epsilon$ Indi Ba,Bb&$-0.014\pm0.003$& $-0.24^{+0.08}_{-0.09}$&$-0.26^{+0.10}_{-0.09}$\\
    \hline
    \end{tabular}
    \caption{Table showing the Pearson (PCC) and Spearman (SCC) correlation coefficients and best-fit slope for the photometry against FWHM (as a proxy for seeing) on the night of 2017-08-28 for the target, and two comparison stars.}
    \label{tab:FWHMCCtable20170820}
\end{table*}
}



\begin{figure}
    \centering
    \includegraphics[width=0.95\columnwidth]{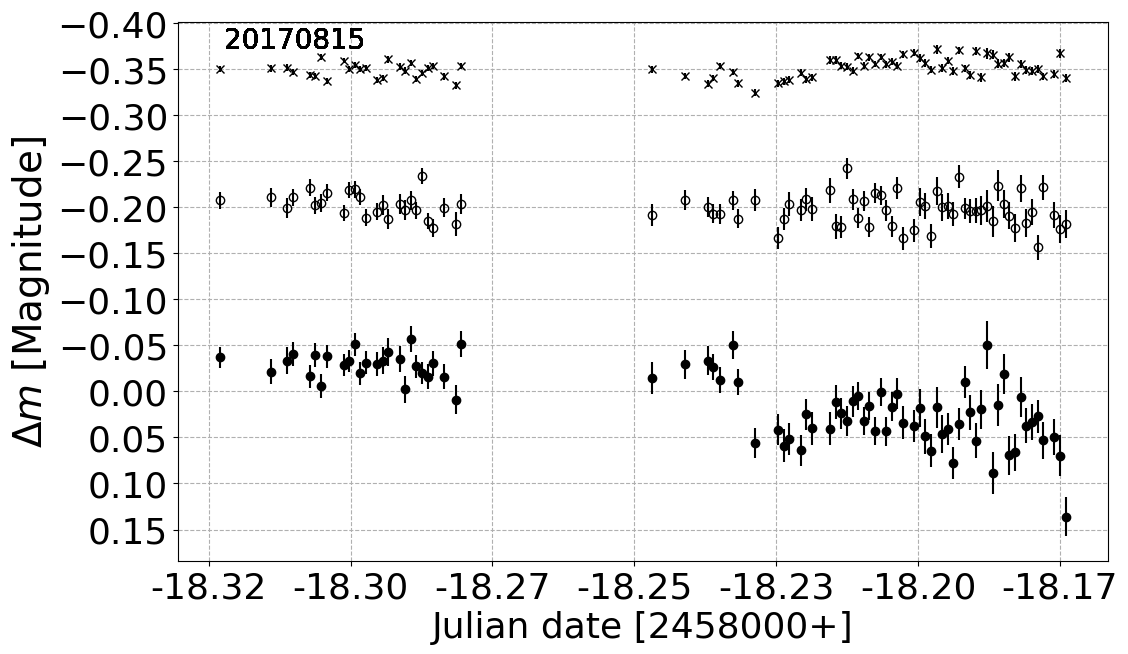}
    \caption{Normalised, arbritrarily offset differential photometry for $\epsilon$ Indi Ba,Bb (filled circles), a red comparison star (hollow circles) and a bright comparison star (crosses) on the night of 2017-08-15. Both intermittent cloud coverage and suspected light contamination from $\epsilon$ Indi A produced spurious photometry on this night, causing the gap in coverage. For these reasons, the 0.05 mag drop in brightness of the target is assumed to not be real.}
    \label{fig:20170815}
\end{figure}

In addition to the night of 2017-08-28, the only other significant within-night variability was seen on the night of 2017-08-15, shown in Figure \ref{fig:20170815}, but this sudden drop of 0.05 mag is suspect. As discussed in Section \ref{sec:comparstar}, both intervening cloud and suspected light contamination by $\epsilon$ Indi A led to spurious photometry during this night, resulting in a $\sim1$ hr interval where there's a gap in the time series, and in the window immediately following this -- where we see a sudden drop in brightness of the target -- similar behaviour was seen in a number of the comparison star time series.

\subsection{Search for lightning}

No clear short-timescale signatures indicative of stochastic, \lq lightning-like\rq\space activity are apparent in the \textit{I}-band time series for all nights (\S \ref{sec:nolightning}). Additionally, 234 \textit{V}-band and 412 H\textit{$\alpha$} images were fed through {\sc Source Extractor} -- which was configured with a $>3\sigma$ above background detection threshold -- to directly search for strong emission signatures of lightning free from the continuum spectrum of the target. Despite two false positives easily identified as cosmic ray hits, no detections at the position of the target were found in these images.

\subsection{Night-to-night variation}

\begin{figure}
    \centering
    \includegraphics[width=9cm,height=9cm,keepaspectratio]{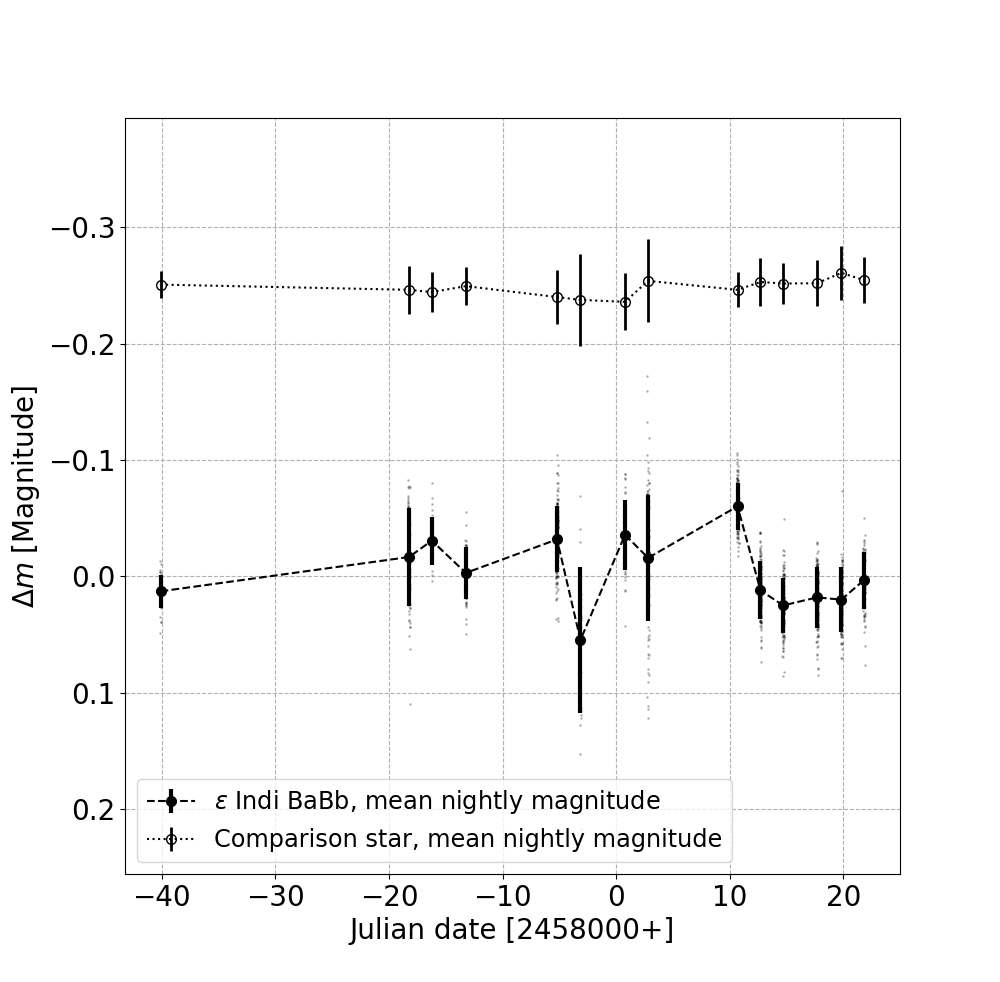}
    \caption{Zero-point corrected differential photometry for the entire observing campaign for the target. Individual differential magnitudes are shown as faint dots, and the mean magnitude for both the target (filled circles) and red comparison star (hollow circles) are plotted as the large markers, with associated error bars. These error bars are the root sum of squares of the standard deviation of the photometry on each night and the error associated with the photometric zero-point shift.}
    \label{fig:allnights+2compar}
\end{figure}

$\epsilon$ Indi Ba,Bb shows large flux variability over the course of the entire 2 months of observations. The full zero-point shifted light curve of the target is shown on the bottom row of Figure \ref{fig:allnights+2compar}. Therein, we plot both the individual differential magnitudes as the small shaded points, in addition to the mean magnitude and its associated error. The errorbars are the root sum of squares of the standard deviation of the photometry on each night and the error associated with the approximation to the photometric zero-point for each night (see Equation \ref{eq:zp}). For reference, the mean magnitudes of a red comparison star - the same one shown in Figures \ref{fig:epsilon_timeseries} and \ref{fig:20170815} - are also plotted.

This red comparison star is both one of the reddest stars in the field and of a similar brightness to the target. That it shows far more stable behaviour in Figure \ref{fig:allnights+2compar} than the target is supportive of the reality of the variation in the target. Indeed, as described in Section \ref{sec:zpcalc}, there was no clear trend of zero-point with source colour for any of the comparison stars.

$\epsilon$ Indi Ba,Bb spans a range of 0.10 mag over the whole campaign. There is a $\sim0.1$ mag increase in brightness over just 4 days, consistent with the previously reported variability in the \textit{I}-band described in Section \ref{sec:intro}. There's no clearly consistent timescale for this large variation. For example, after JD [2458000+] 10 the target fades by $\sim0.07$ mag rapidly over just 2 days, which is immediately followed by a period of relative stability over the next 5 nights covering the last 9 days of the campaign.

Since colour effects are non-linear, it is possible that even small variations in night-to-night systematics may cause large changes in the mean nightly photometry of this very red target. As pointed out in Section \ref{sec:comparstar}, the photometry on the night of 2017-08-30 was affected by changing cloud coverage, and the mean brightness of several field stars was also seen to vary in a similar manner to the target on this night. If one excludes the nights where clouds are known to have affected the observations (i.e. the nights of 2017-08-15, 2017-08-20 and 2017-08-30), we still measure significant variability on a timescale of at least two days, although the net change in brightness over the entire campaign is slightly reduced, at $\sim 0.09$ mag.

\begin{figure}
    \centering
    \includegraphics[width=\columnwidth]{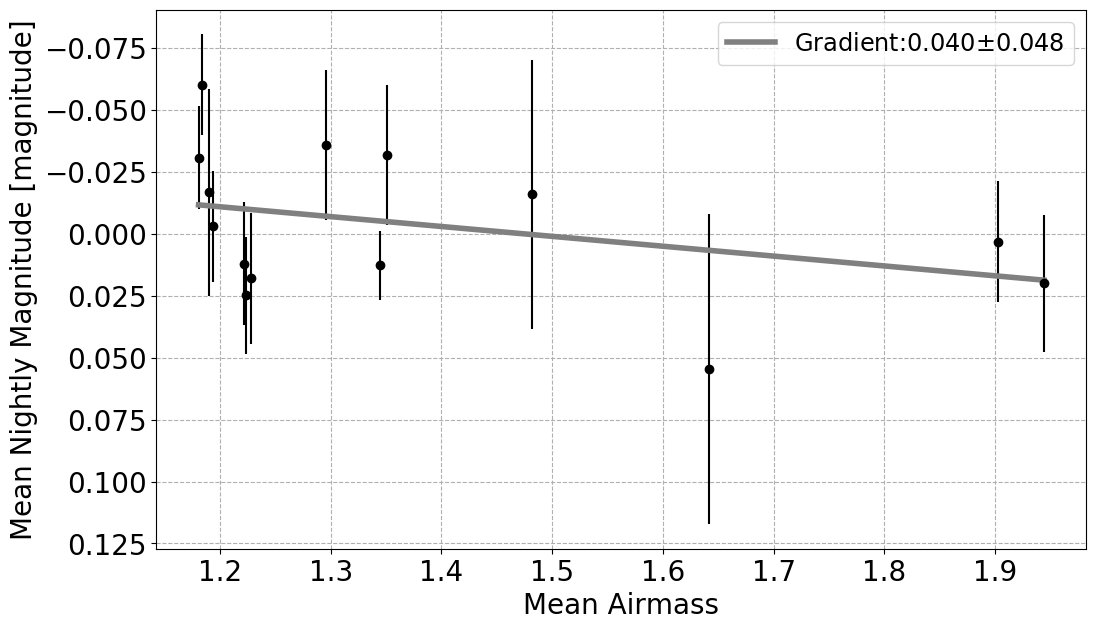}
    \caption{Mean airmass vs. the mean value of the normalised differential photometry of $\epsilon$ Indi Ba,Bb for each night.}
    \label{fig:Airmassallnights}
\end{figure}

\begin{figure}
    \centering
    \includegraphics[width=\columnwidth]{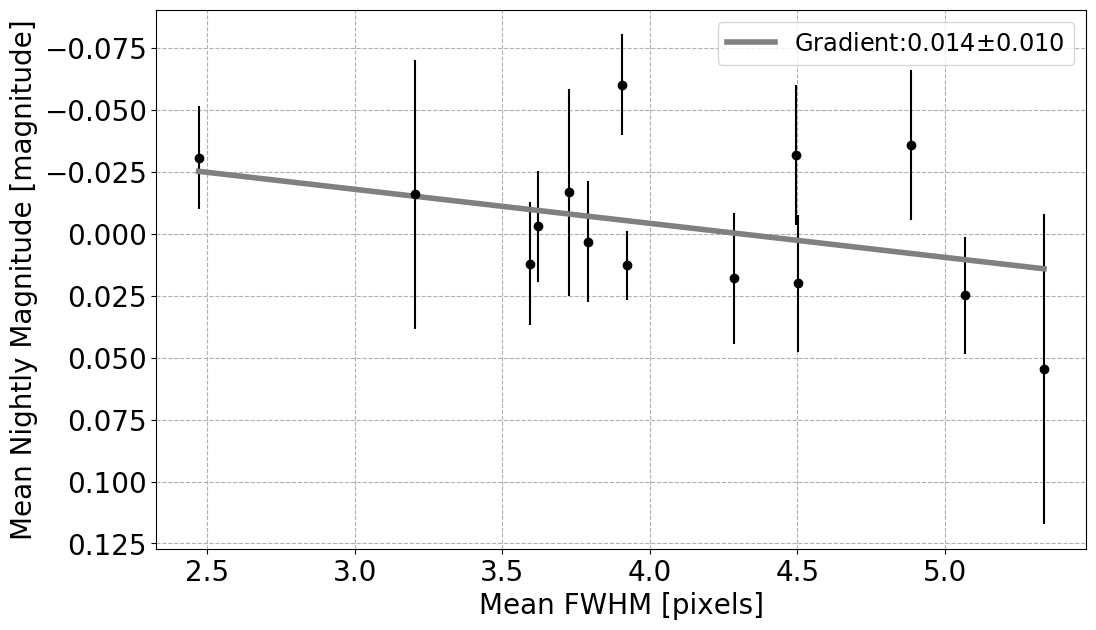}
    \caption{Mean image FWHM (as a proxy for seeing) vs. the mean value of the normalised differential photometry of $\epsilon$ Indi Ba,Bb for each night.}
    \label{fig:Seeingallnights}
\end{figure}

\textcolor{black}{As in Section \ref{sec:withinnight}, we check whether the mean variation in magnitude of the target is correlated with airmass and/or seeing by plotting the mean zero-point shifted magnitude for each night against the corresponding mean values of airmass and FWHM in Figures \ref{fig:Airmassallnights} and \ref{fig:Seeingallnights} respectively. Use of either the Pearson or Spearman correlation coefficients for such small samples is a biased estimate of correlation, and can give spurious results. We see however that there is a large relative uncertainty on the gradient of the best fit line for both plots, which is again estimated empirically by the bootstrap method with 1000 trials (see Equation \ref{eq:bootstrap}). In Figure \ref{fig:Airmassallnights}, the uncertainty exceeds the gradient itself and so is consistent with a flat line. For Figure \ref{fig:Seeingallnights}, although significantly different from a flat line, the relative error is very large, and the gradient itself is far lower than the level of variation we claim to see on this night-to-night timescale. This suggests that the mean brightness of the target is not a strong function of either seeing of airmass.}

\textcolor{black}{Given that there were changes in pointing over the 2 months of observing, it is necessary to include some assessment of any consequent systematic impact on the night-to-night time series. It is expected that this effect, if present, should be similar for neighbouring stars on the chip. For this reason, the quantitative assesmment of night-to-night variaiblity which follows is done with reference to the time series of stars neighbouring the target. Specifically, we select stars that are within a $6\times 6$ arcminute box centered on the target.}

\begin{figure}
    \centering
    \includegraphics[width=9cm,height=9cm,keepaspectratio]{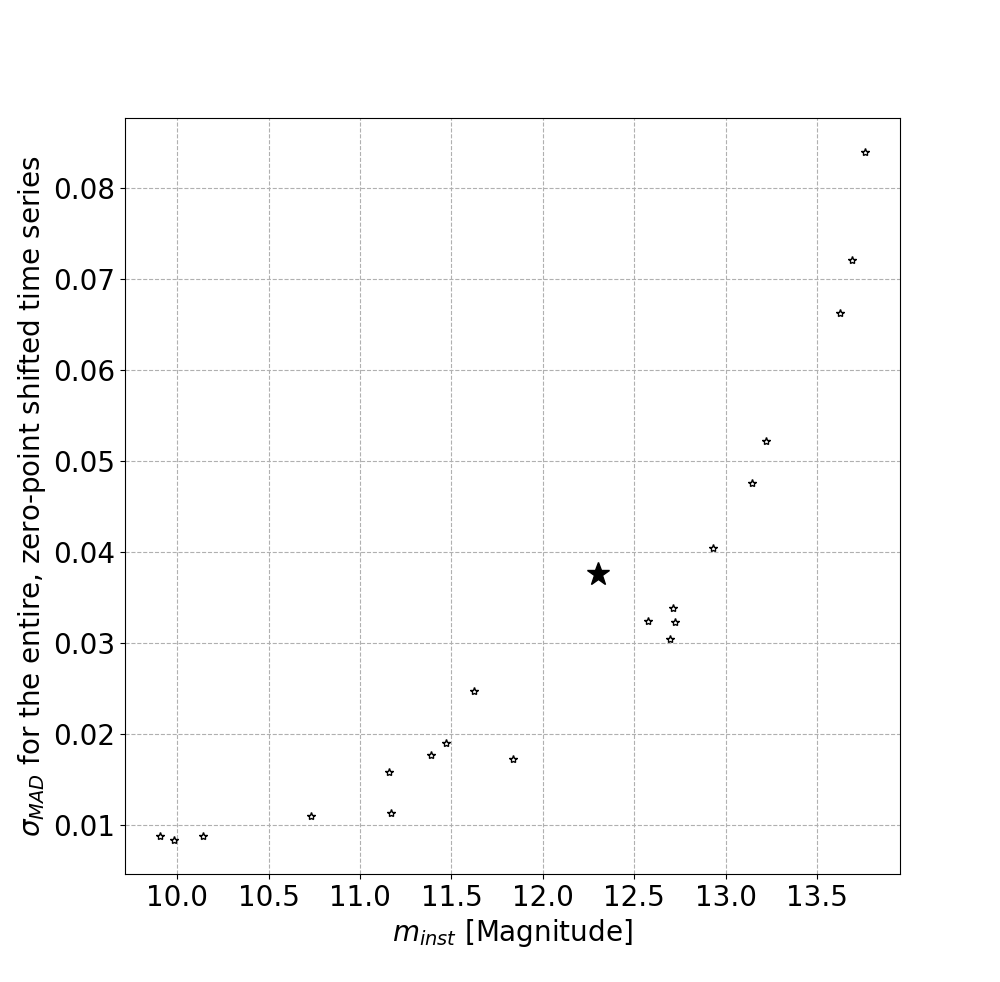}
    \caption{The median absolute deviation -- scaled to a standard deviation -- of the entire zero-point shifted times series vs instrumental magnitude for stars (small hollow stars) near the target (large filled black star) on the chip.}
    \label{fig:allnightlocal}
\end{figure}

\textcolor{black}{We plot the instrumental mangitude against the $\sigma_{MAD}$ of the entire zero-point corrected $\sim$2 month time series in Figure \ref{fig:allnightlocal} for the target (large filled black star) and its nearest neighbours (small hollow stars). Certainly, this suggests enhanced variability in the target over the 2 months. To quantify how the time series in Figure \ref{fig:allnights+2compar} differs from the simple model of a straight line (with Gaussian measurement noise), one can generate a histogram of the normalised residuals relative to the mean magnitude level, $\overline{m}_{\textrm{inst}}$, which for any magnitude measurement $m_{\textrm{inst}, n}$ with uncertainty $\sigma_n$ is,}

\begin{equation}
    R_n = \frac{m_{\textrm{inst}, n} - \overline{m}_{\textrm{inst}}}{\sigma_n}
    \label{eq:normresid}
\end{equation}

\begin{figure}
    \centering
    \includegraphics[width=8cm,height=9cm,keepaspectratio]{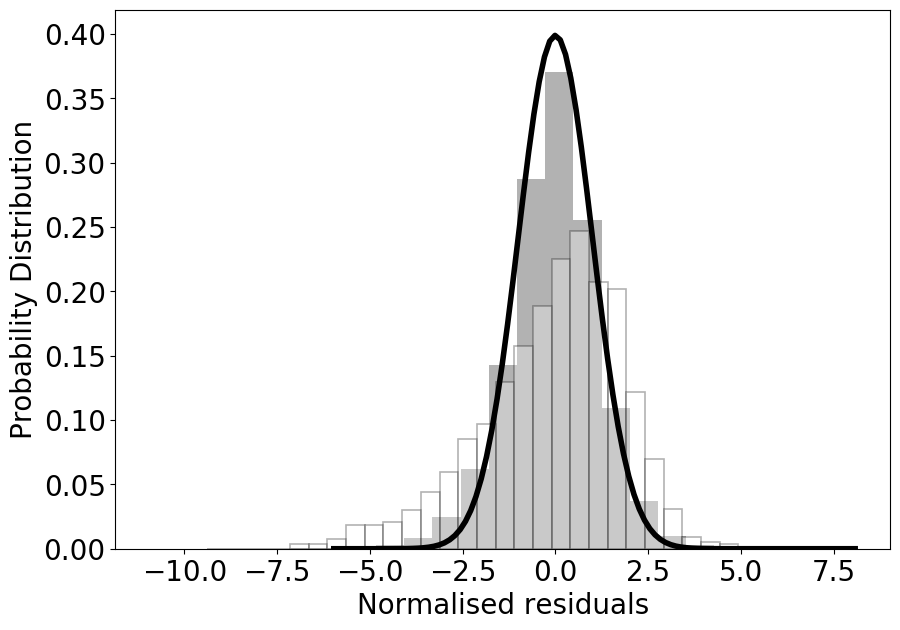}
    \caption{Normalised histograms for the normalised residuals of the target (white) and neighbouring stars, binned collectively into an empirical distribution (grey). A unit Gaussian is overlain for comparison, and shows clear similarities with the empirical distribution. This constrasts with the target's histogram, which is statistically signficantly different than the empirical distribution (p-value from KS test is $3.1\times 10^{-24}$) and shows clear asymmetry.}
    \label{fig:histempirical}
\end{figure}

\textcolor{black}{and compare the normalised histogram -- such that the counts integrate to 1 -- against a Gaussian with mean 0 and variance 1. A Kolmogorov-Smirnoff test can then be used to compare the similarity of the two distributions \citep{massey1951kolmogorov}. Given that systematics in the photometry could introduce non-Gaussian noise into the night-to-night time series, a better approach is to compare the normalised residuals of the target time series against an \textit{empirical} distribution. To construct this, we generate a histogram of the normalised residuals of the neighbouring stars -- the same as in Figure \ref{fig:allnightlocal} -- and perform a KS test between this empirical distribution, and that of the target. This gives us a quantitative measure of how different the target's night-to-night variation is from the neighbouring stars. We show this in Figure \ref{fig:histempirical}, where the target's distribution (white histogram) is overlain onto this empirical distribution (grey). By-eye inspection shows clear differences between the two distributions -- note, for example, the asymmetry in the target's histogram -- and the returned p-value from a KS test is very low, at $3.1\times 10^{-24}$.  A unit Gaussian is overlain for comparison, which as expected, shows similarities with the empirical distribution, but clearly deviates from the target's.}

\section{Discussion}\label{sec:discussion}

\subsection{No photometric period}

The rotation periods of $\epsilon$ Indi Ba and Bb are not known. Provided the photometric signatures of the rotation periods of each of the brown dwarfs exceed the levels of variation associated with evolving weather systems or flaring (and of course, the inherent white noise) one might expect to see a superposition of two distinct periods. In reality, we expect the T1 component to dominate any photometric signal, due both to its greater brightness, and presumably greater cloud coverage given its evolutionary proximity to the L/T transition. Indeed, from the resolved \textit{I}-band photometry in Table \ref{tab:epsiinfo}, the T6 component only contributes 20 per cent of the total \textit{I}-band flux. If all the measured variation were associated with this fainter component, it would have to be implausibly variable -- by more than 50 per cent -- to reproduce the variation we see. 

Neither a Lomb-Scargle periodogram \citep{scargle1982studies} nor autocorrelation function analysis applied to the individual nightly time series returned a consistent period. A flexible Gaussian Process model described by a periodic kernel was conditioned on the entire data set \citep{foreman2017fast}. The marginal posterior distribution of the period parameter was sampled with a Markov Chain Monte Carlo method, and this too failed to reveal any convincing periodicity.

\subsection{No evidence for lightning}\label{sec:nolightning}
\subsubsection{\textcolor{black}{A test for asymmetry in the within-night time series}}

Following \citet{cody2014csi}, for each night of \textit{I}-band observations, we define the metric $M$, such that

\begin{equation}
    M = (\langle d_{10\%} \rangle - d_{\text{med}})/\sigma_d ,
    \label{eq:M_metric}
\end{equation}

where $\langle d_{10\%} \rangle$ is the mean of the $10\%$ highest and lowest magnitude values of the time series, and $d_{\text{med}}$ and $\sigma_d$ are the median flux level and rms scatter for the entire nightly time series. In all other time series analyses presented in this work (i.e. the period search, monitoring of the long-term changes in mean brightness etc.) a $4\sigma$ clip was used for the differential photometry. In this subsection only -- which describes an analysis specifically designed to probe for possible lightning strikes -- we slightly relax this constraint, and instead clip $5\sigma$ outliers from the within-night differential photometry. This was done to continue to guard against both spurious photometry and comsic ray hits, but also allow for potentially large spikes in brightness. 

\begin{figure}
    \centering
    \includegraphics[width=\columnwidth]{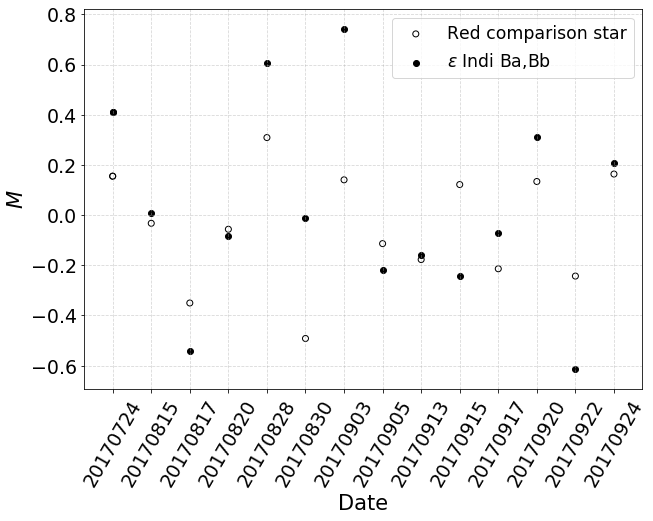}
    \caption{Values for the metric described by Equation \ref{eq:M_metric} on each night, for both the target (filled circles) and a red comparison star (hollow circles).}
    \label{fig:M_metric}
\end{figure}

Application of Equation \ref{eq:M_metric} to each of the individual nightly $\epsilon$ Indi Ba,Bb time series revealed no consistent asymmetry above the median flux level. For lightning-like signatures (i.e. a stochastic brightening above the median flux level), one would expect to recover negative values of $M$. Rather, both positive and negative values for $M$ were found, randomly distributed, with a mean and median of 0.025 and -0.043 (see Figure \ref{fig:M_metric}). The largest absolute values of $M$ were seen to correspond to nights with occasional \lq blips\rq\space above or below the median flux level, or more rarely, on nights with a smooth, weakly underlying trend (e.g. 2017-08-28). The \lq blips\rq\space in the target's time series are also seen on occasion in the comparison stars' within-night time series, producing similar values of $M$, which suggests imperfections with the photometry, and not any real variation. Indeed, the $M$ values for the red comparison star whose time series are shown alongside the target in Figure \ref{fig:epsilon_timeseries} are also plotted, and show a similar spread in $M$ values, albeit at a somewhat lower amplitude.  The sensitivity of the $M$ metric to these systematic \lq blips\rq\space in an otherwise quiet time series is not surprising, as its value  will be strongly influenced by these few, outlying points. Indeed, in contrast to the time series analysed here, \citet{cody2014csi} apply this metric to time series of clearly variable sources, where the level of true astrophysical variation greatly exceeds the noise.

\subsubsection{\textcolor{black}{Implications for the properties of extrasolar lightning}}

One could consider what upper limits these non-detections for lightning -- both in the \textit{I}-band time series and the \textit{V} and H\textit{$\alpha$} images -- place on the physical properties of the strikes in the atmospheres of brown dwarfs. However, the physical parameters which describe the properties of lightning in extrasolar environments remain difficult to constrain given our non-detections. These include the power and duration of strikes, their flash density (i.e. the rate of strikes per unit area), percentage coverage of strikes over the brown dwarf's hemisphere and how the power is radiated into frequencies across the optical (see e.g. \citet{bailey2014ionization}, \citet{hodosan2016lightning}). \textcolor{black}{Consequently, all we can confidently say, is that (1) if lightning strikes on $\epsilon$ Indi Ba,Bb are indeed observable above the noise in our \textit{I}-band time series (with median $\sigma_{\text{MAD}} = 21$ mmag) we do not observe any strikes in a total $\sim 42$ hours of coverage. That is, the rate of lightning strikes that emerge out of the atmosphere is $< 0.02$ strikes per hour. (2) We do not see anything like the increase in brightness estimated from the most promising \textit{combination} of parameters detailed in \citet{hodosan2017lightning}.}

\subsection{Days-long variability}

\textcolor{black}{The lack of any clear periodicity and significant days-long variability could suggest we are viewing the system close to a pole. The large changes in mean brightness suggest inhomegenity of surface features on the brighter T1 component. Given the extreme coolness of the target (see Table \ref{tab:epsiinfo}), these almost certainly arise from large-scale changes in the cloud coverage. By large-scale, we simply mean that the cause of the variation is not localised to any single part of the surface i.e. the variation is caused by some net effect due to the changing cloud configurations expected to cover this ultracool target. The lack of any other clear signal can be viewed as beneficial, in that we measure the \lq pure\rq\space signal associated with these inhomogeneities only, free from any rotational influence. Although cloud coverage is the most likely explanation for the variability, we note that the plausibility of any particular scenario will be dependent on spectroscopic follow-up of this and other objects, in addition to numerous theoretical assessments.}

If large-scale changes in the cloud coverage are responsible for changes in mean brightness -- which is active on time-scales as short as 2 days -- these clouds might occur with a banded structure similar to the striking clouds seen in Jupiter's atmosphere \citep{apai2017zones}. The regions of cyclonic shear at the boundaries between these bands are a perfect environment for generating lightning, as seen for Jupiter and Saturn in our own solar system \citep{little1999galileo}. Our current inability to put tight prior constraints on the expected properties of lightning makes planning searches designed to probe these signatures far from straightforward.

Certainly, this work suggests that conventional CCD imaging in either narrow or broadband filters is not an ideal strategy to detect such signatures, even at a fairly high cadence. Given the expected sub-second duration of any individual lightning strike, the shorter the exposure time, the more easily detectable a given strike will be above the background flux of the hosting source. Technological advances with high-frame rate cameras -- which make use of special low-light-level detectors (e.g. Electron-multiplying CCDs, and more recently, CMOS Imaging Sensors) -- have extended time domain astronomy to this sub-second regime. The simultaneous multi-wavelength, high-cadence studies made possible by attaching these devices to medium to large-sized telescopes (e.g. the recent OPTICAM instrument \citep{castro2019opticam}) may provide the crucial observational requirements.


\section{Conclusions}\label{sec:conclusion}

We have discussed the results of an analysis of 14 nights of \textit{I}-band photometric monitoring of the nearby $\epsilon$ Indi Ba,Bb brown dwarf binary. The target typically appears to be unremarkably quiet on the hour-long timescales of each night of observing, which contrasts with the large changes in mean brightness - by as much as $0.10$ mag - which we measure between nights across the entire 2 month campaign. The hours-long timescale stability of the target, and lack of any clear periodicity, suggests that the large changes in mean brightness may arise from changes in the large-scale cloud structure. The regular nightly visits to this target and overall long-term coverage of this data set provides a new insight into the irregular, days-long variability exhibited by brown dwarfs at the L/T transition. Indeed, we expect this signal is associated with $\epsilon$ Indi Ba, which is both the brighter of the two components, and of spectral type T1, and so likely far cloudier than its cooler T6 companion.

Lightning will very likely occur where dynamic clouds form, and complementary \textit{V}-band and H\textit{$\alpha$} images were acquired to search for stochastic signatures diagnostic of the lightning strikes assumed to be present in the atmospheres of these brown dwarfs. No $>3\sigma$ detections above the background at the position of the target were found in these passbands, nor is there any suggestion of short-timescale, asymmetric \lq flickering\rq\space in the \textit{I}-band time series. The necessarily long exposures required for conventional CCDs -- such that the signal of interest exceeds the readout noise -- limits our ability to detect these short-timescale events. Exploration of the sub-second time variability of astrophysical sources is being made increasingly feasible with technological advances of high frame-rate detectors, and we recommend future lightning hunters adopt these new technologies for their searches.

\section*{Acknowledgements}

We extend a collective thank you to the MiNDSTEp consortium -- both the observers and those responsible for the smooth running of operations -- for adopting this side project for the 2017 observing season. L.M. acknowledges support from the University of Rome Tor Vergata through ``Mission: Sustainability 2017'' fund. \ch{ChH and GH  gratefully acknowledge the support of the ERC
Starting Grant no. 257431.}




\bibliographystyle{mnras}
\bibliography{epsindi} 




\appendix

\section{Supplementary time series}\label{sec:supptimeseries}

\begin{figure*}
    \centering
    \subfloat{
        \centering
        \includegraphics[width=0.48\textwidth, height=6cm]{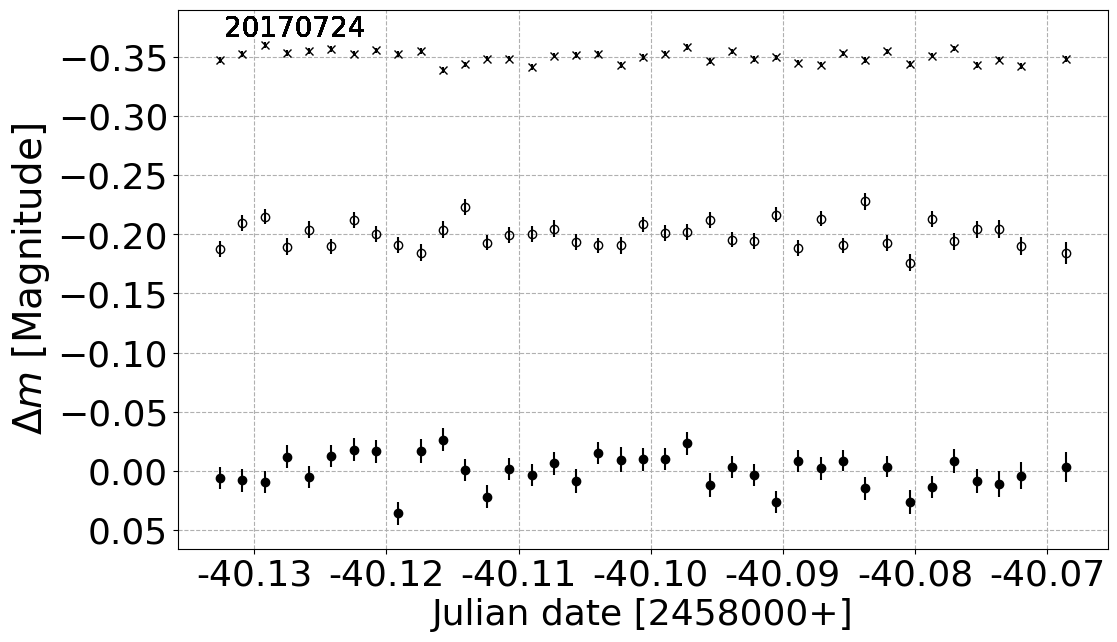}
    }
    \hfill
    \subfloat{
        \centering 
        \includegraphics[width=0.48\textwidth, height=6cm]{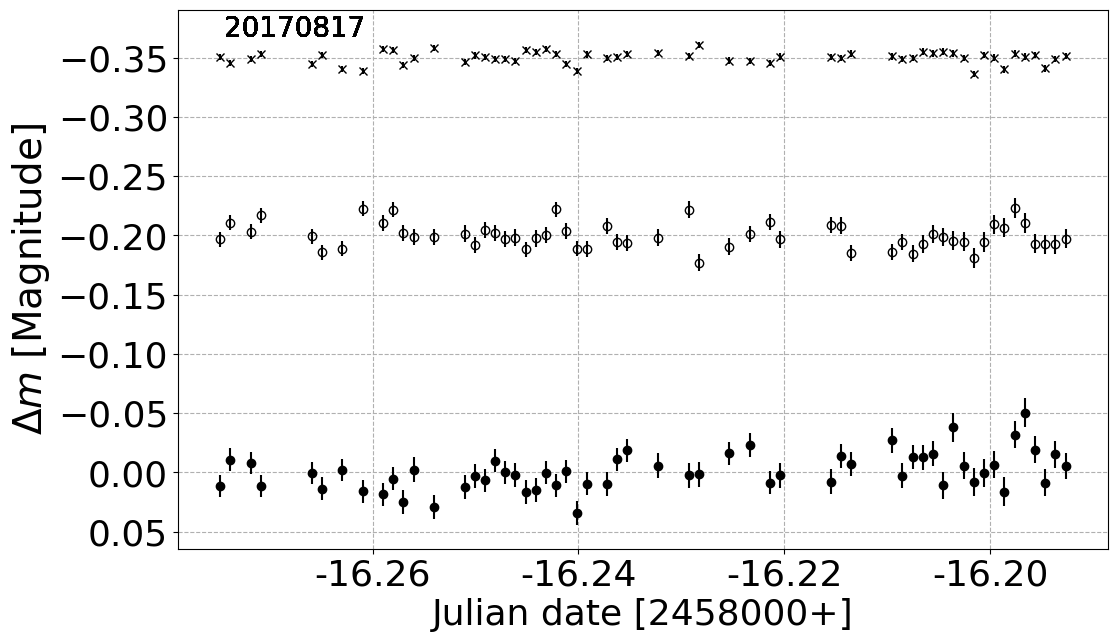}
    }
    \vskip\baselineskip
    \subfloat{
        \centering
        \includegraphics[width=0.48\textwidth, height=6cm]{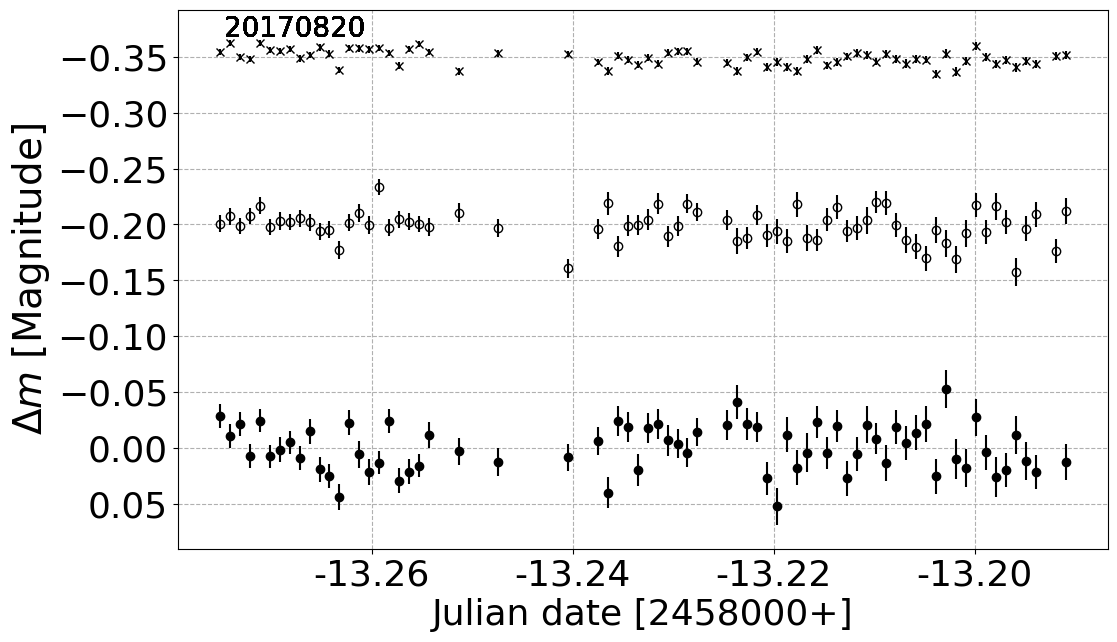}
    }
    \hfill
    \subfloat{ 
        \centering 
        \includegraphics[width=0.48\textwidth, height=6cm]{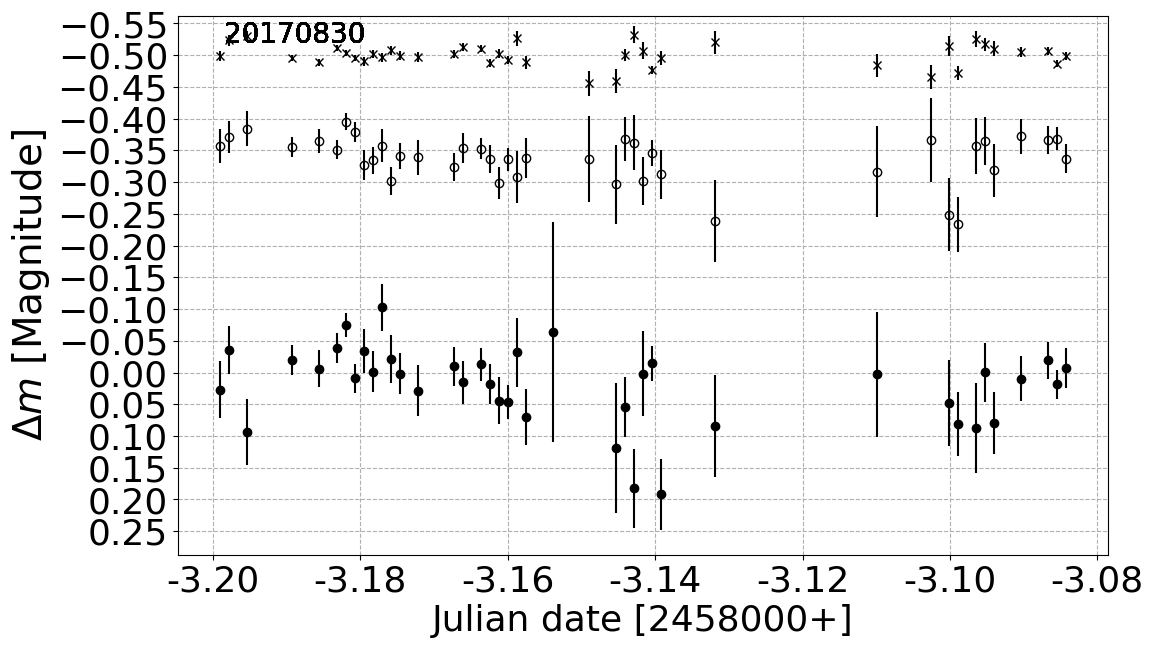}
    }
    \vskip\baselineskip
    \subfloat{
        \centering 
        \includegraphics[width=0.48\textwidth, height=6cm]{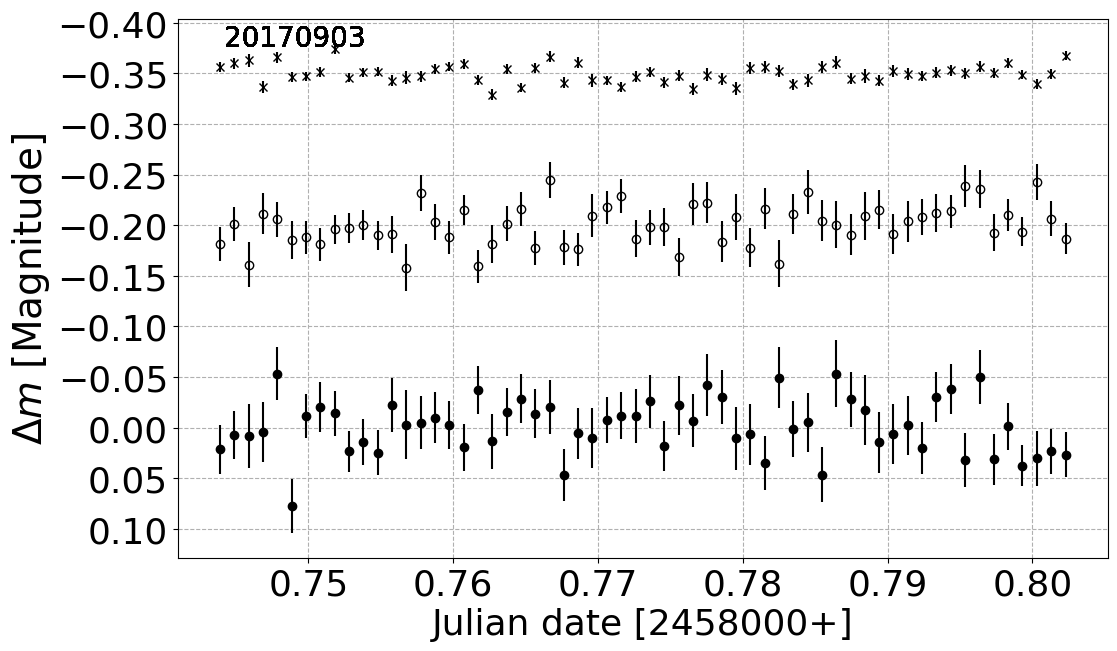}
    }
    \quad
    \subfloat{   
        \centering 
        \includegraphics[width=0.48\textwidth, height=6cm]{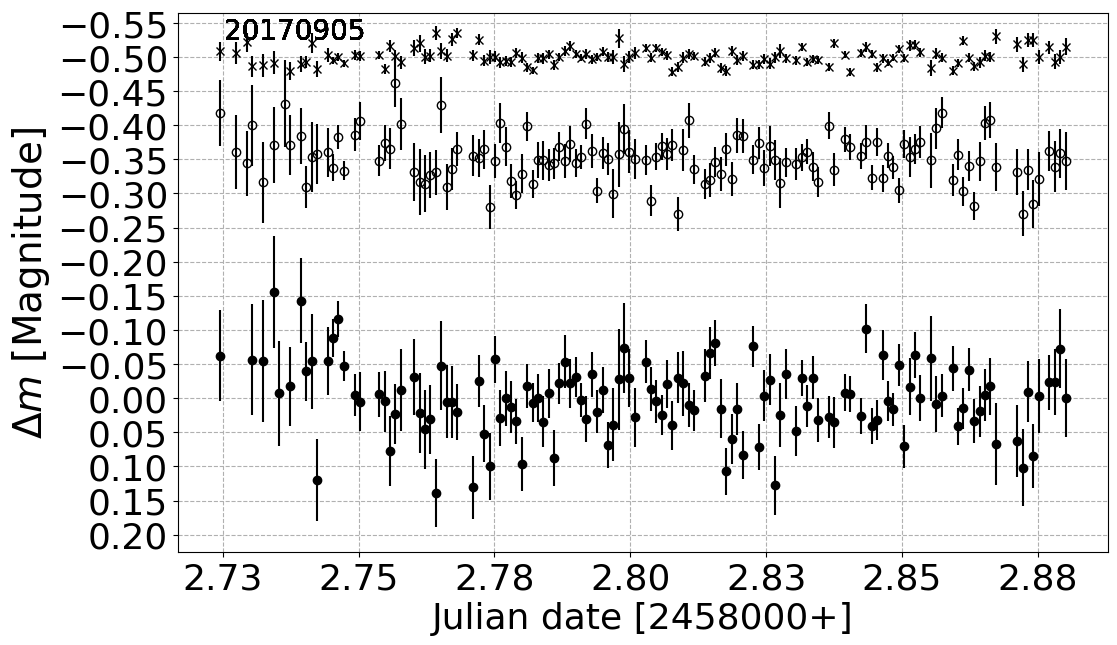}
    }
    \caption{Normalised differential photometry, as in Figure \ref{fig:epsilon_timeseries}, for 6 additional nights not shown in Section \ref{sec:genvariability}. As noted in Section \ref{sec:obs+datared}, observations on the night of 2017-08-30 were interrupted by passing clouds.}
    \label{fig:epsilon_timeseries_a1}
\end{figure*}

\begin{figure*}
    \centering
    \subfloat{
        \centering 
        \includegraphics[width=0.48\textwidth, height=6cm]{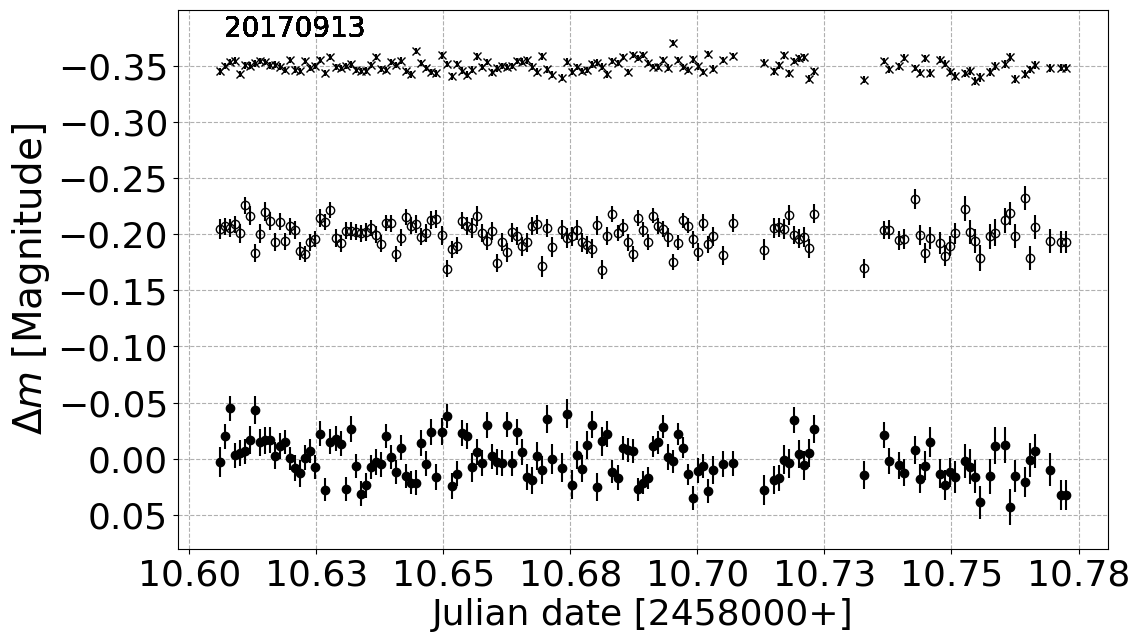}
    }
    \quad
    \subfloat{
        \centering 
        \includegraphics[width=0.48\textwidth, height=6cm]{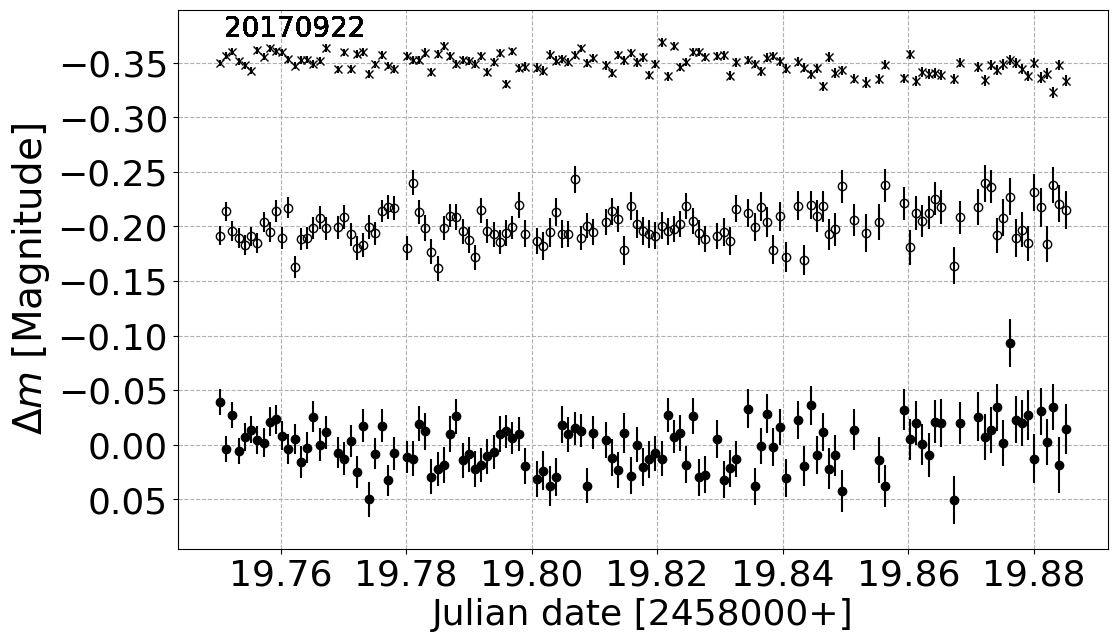}
    }
    \vskip\baselineskip
    \subfloat{
        \centering 
        \includegraphics[width=0.48\textwidth, height=6cm]{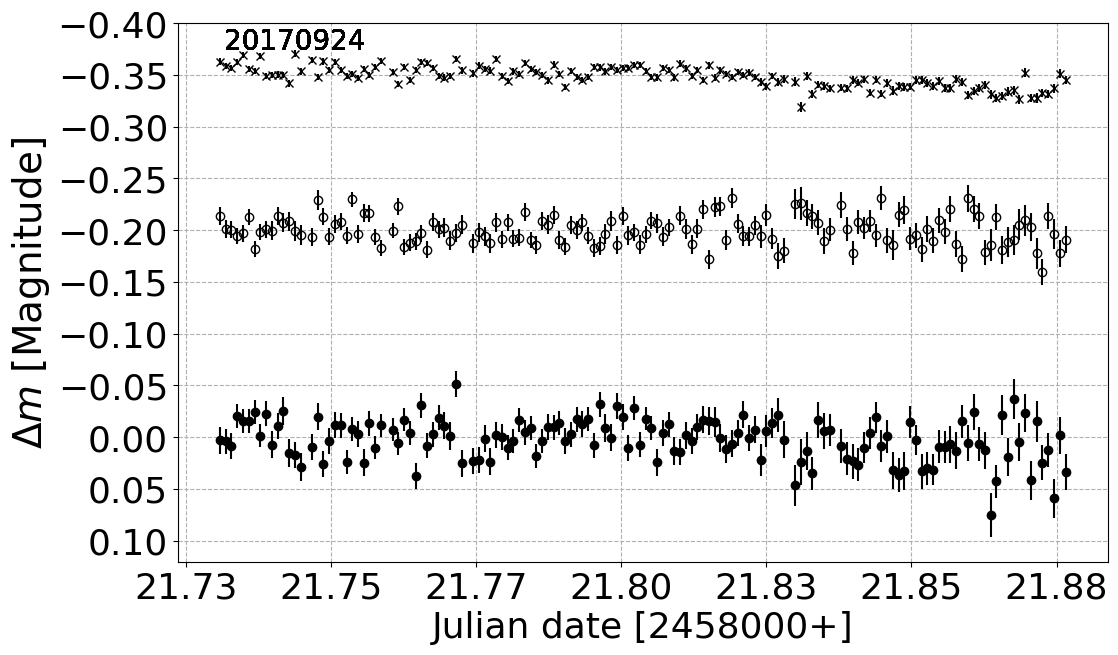}
    }
    \caption{Normalised differential photometry, as in Figure \ref{fig:epsilon_timeseries}, for 3 additional nights not shown in Section \ref{sec:genvariability}.}
    \label{fig:epsilon_timeseries_a2}
\end{figure*}

The normalised differential photometry for the remaining 9 nights which are not shown in Section \ref{sec:genvariability} are plotted in Figures \ref{fig:epsilon_timeseries_a1} and \ref{fig:epsilon_timeseries_a2}.

The \textit{I}-band differential photometry for all nights is publicly available for download in machine-readable form via ScholarOne. We include the target time series, and those of the 16-large comparison star ensemble to allow independent validation of the night-to-night zero-point calculations. These instrumental magnitudes have not been transformed to a standard photometric system, and are set on an arbitrary scale. The first 5 rows are shown in Table \ref{tab:bigtable}.

\begin{table*}
	\centering
	\label{tab:example_table}
	\caption{An extract of the first 5 rows of the table hosting the \textit{I}-band differential photometry used in this work for the target and comparison stars. These instrumental magnitudes have not been transformed to a standard photometric system, and are set on an arbitrary scale. Here, the photometry measured on different nights has not been zero-point corrected. \textcolor{black}{The full table is available in machine-readable form online.}}
	\begin{tabular}{cccccccccc}
		\hline
		\begin{tabular}{@{}c@{}}$\epsilon$ Indi Ba, Bb \\ JD [2458000+] \end{tabular} & $m_{\mathrm{inst}}$ & $m_{\mathrm{err}}$ & \begin{tabular}{@{}c@{}} Comparison1 \\ JD [2458000+] \end{tabular} & $m_{\mathrm{inst}}$ & $m_{\mathrm{err}}$ & ... & \begin{tabular}{@{}c@{}} Comparison16 \\ JD [2458000+] \end{tabular} & $m_{\mathrm{inst}}$ & $m_{\mathrm{err}}$ \\
        \hline
        -40.1326 & 12.3171 & 0.0090 & -40.1326 & 11.5945 & 0.0073 & ... & -40.1326 & 11.3186 & 0.0069 \\
        -40.1309 & 12.3193 & 0.0097 & -40.1309 & 11.6133 & 0.0077 & ... & -40.1309 & 11.3411 & 0.0071  \\
        -40.1292 & 12.3207 & 0.0093 & -40.1292 & 11.6081 & 0.0075 & ... & -40.1292 & 11.3319 & 0.0072 \\
        -40.1275 & 12.2992 & 0.0097 & -40.1275 & 11.6029 & 0.0080 & ... & -40.1275 & 11.3228 & 0.0075 \\
        -40.1258 & 12.3165 & 0.0093 & -40.1258 & 11.6110 & 0.0073 & ... & -40.1258 & 11.3291 & 0.0069 \\
        ... & ... & ... & ... & ... & ... & ... & ... & ... & ... \\
        \hline

	\end{tabular}
	\label{tab:bigtable}
\end{table*}

\section{Supplementary correlation plots}\label{sec:suppcorrplots}

{ 
\renewcommand{\arraystretch}{1.5}
\begin{table*}
    \centering
    \begin{tabular}{c|c|c|c}
    \hline
    Date & Star & Slope [vs Airmass] & SCC [vs Airmass]\\
    \hline
    2017-07-24 & Red comparison star&$0.007\pm0.028$& $0.05^{+0.18}_{-0.19}$ \\
    
    & $\epsilon$ Indi Ba,Bb&$0.034\pm0.030$& $0.14^{+0.19}_{-0.18}$\\
    
    2017-08-15 & Red comparison star&$0.137\pm0.122$& $0.04^{+0.14}_{-0.14}$\\
    
    & $\epsilon$ Indi Ba,Bb&$0.588\pm0.054$& $0.65^{+0.07}_{-0.08}$\\
    
    2017-08-17 & Red comparison star&$0.056\pm0.039$& $0.13^{+0.15}_{-0.16}$\\
    
    & $\epsilon$ Indi Ba,Bb&$-0.193\pm0.050$& $-0.37^{+0.15}_{-0.12}$\\
    
    2017-08-20 & Red comparison star&$0.062\pm0.036$& $0.16^{+0.14}_{-0.14}$\\
    
    & $\epsilon$ Indi Ba,Bb&$-0.011\pm0.058$& $0.03^{+0.14}_{-0.14}$\\
    
    2017-08-30 & Red comparison star&$0.010\pm0.021$& $0.17^{+0.21}_{-0.20}$\\
    
    & $\epsilon$ Indi Ba,Bb&$0.044\pm0.030$& $0.28^{+0.16}_{-0.18}$\\
    
    2017-09-03 & Red comparison star&$-0.100\pm0.040$& $-0.25^{+0.15}_{-0.13}$\\
    
    & $\epsilon$ Indi Ba,Bb&$0.046\pm0.065$& $0.02^{+0.17}_{-0.16}$\\
    
    2017-09-05 & Red comparison star&$0.012\pm0.015$& $0.11^{+0.11}_{-0.12}$\\
    
    & $\epsilon$ Indi Ba,Bb&$0.027\pm0.023$&$0.09^{+0.12}_{-0.11}$\\
    
    2017-09-13 & Red comparison star&$0.009\pm0.015$& $0.04^{+0.12}_{-0.10}$\\
    
    & $\epsilon$ Indi Ba,Bb&$0.075\pm0.020$& $0.18^{+0.13}_{-0.08}$\\
    
    2017-09-15 & Red comparison star&$-0.036\pm0.013$& $-0.20^{+0.09}_{-0.08}$\\
    
    & $\epsilon$ Indi Ba,Bb&$0.061\pm0.016$& $0.22^{+0.08}_{-0.09}$\\
    
    2017-09-17 & Red comparison star&$-0.006\pm0.011$& $0.00^{+0.11}_{-0.13}$\\
    
    & $\epsilon$ Indi Ba,Bb&$-0.033\pm0.017$& $-0.11^{+0.10}_{-0.09}$\\
    
    2017-09-20 & Red comparison star&$-0.015\pm0.010$& $-0.07^{+0.08}_{-0.09}$\\
    
    & $\epsilon$ Indi Ba,Bb&$0.074\pm0.017$& $0.21^{+0.12}_{-0.09}$\\
    
    2017-09-22 & Red comparison star&$-0.012\pm0.003$& $-0.26^{+0.12}_{-0.05}$\\
    
    & $\epsilon$ Indi Ba,Bb&$-0.006\pm0.005$& $-0.09^{+0.11}_{-0.11}$\\
    
    2017-09-24 & Red comparison star&$0.002\pm0.003$& $0.07^{+0.14}_{-0.14}$\\
    
    & $\epsilon$ Indi Ba,Bb&$0.012\pm0.004$& $0.19^{+0.12}_{-0.11}$\\
    \hline
    \end{tabular}
    \caption{Table showing the Spearman correlation coefficient (SCC) and best-fit slope for the photometry against airmass over several nights for the target and a comparison star.}
    \label{tab:AirmassCCtable}
\end{table*}
}

{ 
\renewcommand{\arraystretch}{1.5}
\begin{table*}
    \centering
    \begin{tabular}{c|c|c|c}
    \hline
    Date & Star & Slope [vs FWHM] & SCC [vs FWHM]\\
    \hline
    2017-07-24 & Red comparison star&$0.005\pm0.009$& $0.06^{+0.20}_{-0.19}$\\
    
    & $\epsilon$ Indi Ba,Bb&$-0.008\pm0.011$& $-0.11^{+0.20}_{-0.17}$\\
    
    2017-08-15 & Red comparison star&$0.006\pm0.003$& $0.16^{+0.12}_{-0.13}$\\
    
    & $\epsilon$ Indi Ba,Bb&$0.047\pm0.007$& $0.64^{+0.08}_{-0.09}$\\
    
    2017-08-17 & Red comparison star&$0.007\pm0.009$& $0.05^{+0.16}_{-0.18}$\\
    
    & $\epsilon$ Indi Ba,Bb&$-0.028\pm0.013$& $-0.20^{+0.17}_{-0.15}$\\
    
    2017-08-20 & Red comparison star&$0.009\pm0.004$& $0.25^{+0.12}_{-0.14}$\\
    
    & $\epsilon$ Indi Ba,Bb&$0.001\pm0.006$& $0.01^{+0.14}_{-0.14}$\\
    
    2017-08-30 & Red comparison star&$0.000\pm0.020$& $0.03^{+0.21}_{-0.20}$\\
    
    & $\epsilon$ Indi Ba,Bb&$0.037\pm0.024$& $0.14^{+0.17}_{-0.19}$\\
    
    2017-09-03 & Red comparison star&$-0.002\pm0.004$& $-0.06^{+0.18}_{-0.15}$\\
    
    & $\epsilon$ Indi Ba,Bb&$-0.009\pm0.005$& $-0.13^{+0.17}_{-0.15}$\\
    
    2017-09-05 & Red comparison star&$-0.006\pm0.010$& $0.06^{+0.11}_{-0.11}$\\
    
    & $\epsilon$ Indi Ba,Bb&$-0.006\pm0.010$& $0.06^{+0.11}_{-0.10}$\\
    
    2017-09-13 & Red comparison star&$0.004\pm0.003$& $0.09^{+0.04}_{-0.06}$\\
    
    & $\epsilon$ Indi Ba,Bb&$0.014\pm0.004$& $0.18^{+0.09}_{-0.10}$\\
    
    2017-09-15 & Red comparison star&$0.005\pm0.003$& $0.10^{+0.08}_{-0.09}$\\
    
    & $\epsilon$ Indi Ba,Bb&$0.007\pm0.004$& $0.08^{+0.09}_{-0.09}$\\
    
    2017-09-17 & Red comparison star&$0.002\pm0.004$& $0.02^{+0.08}_{-0.05}$\\
    
    & $\epsilon$ Indi Ba,Bb&$0.000\pm0.005$& $0.01^{+0.10}_{-0.07}$\\
    
    2017-09-20 & Red comparison star&$-0.001\pm0.002$& $-0.01^{+0.09}_{-0.09}$\\
    
    & $\epsilon$ Indi Ba,Bb&$0.017\pm0.003$& $0.38^{+0.07}_{-0.07}$\\
    
    2017-09-22 & Red comparison star&$-0.009\pm0.004$& $-0.15^{+0.09}_{-0.15}$\\
    
    & $\epsilon$ Indi Ba,Bb&$0.002\pm0.006$& $0.04^{+0.11}_{-0.11}$\\
    
    2017-09-24 & Red comparison star&$0.000\pm0.002$& $0.01^{+0.08}_{-0.08}$\\
    
    & $\epsilon$ Indi Ba,Bb&$0.012\pm0.003$& $0.28^{+0.10}_{-0.11}$\\
    \hline
    \end{tabular}
    \caption{Table showing the Spearman correlation coefficient (SCC) and best-fit slope for the photometry against FWHM (as a proxy for seeing) over several nights for the target and a comparison star.}
    \label{tab:FWHMCCtable}
\end{table*}
}

\begin{figure*}
    \centering
    \subfloat{
        \centering
        \includegraphics[width=0.48\textwidth, height=6cm]{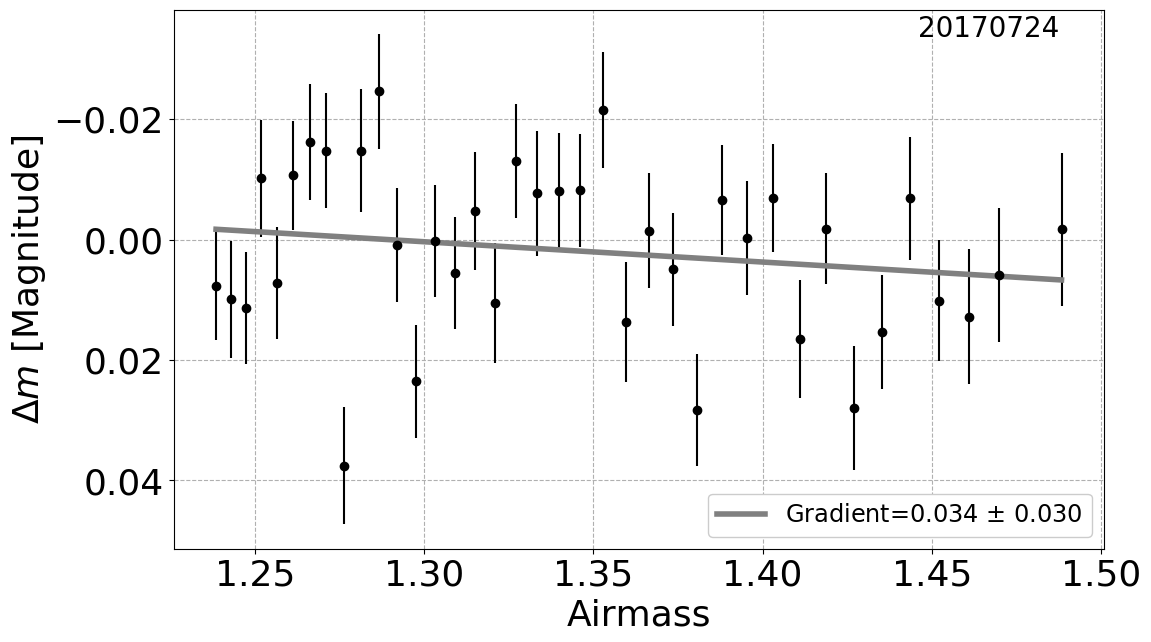}
    }
    \hfill
    \subfloat{
        \centering 
        \includegraphics[width=0.48\textwidth, height=6cm]{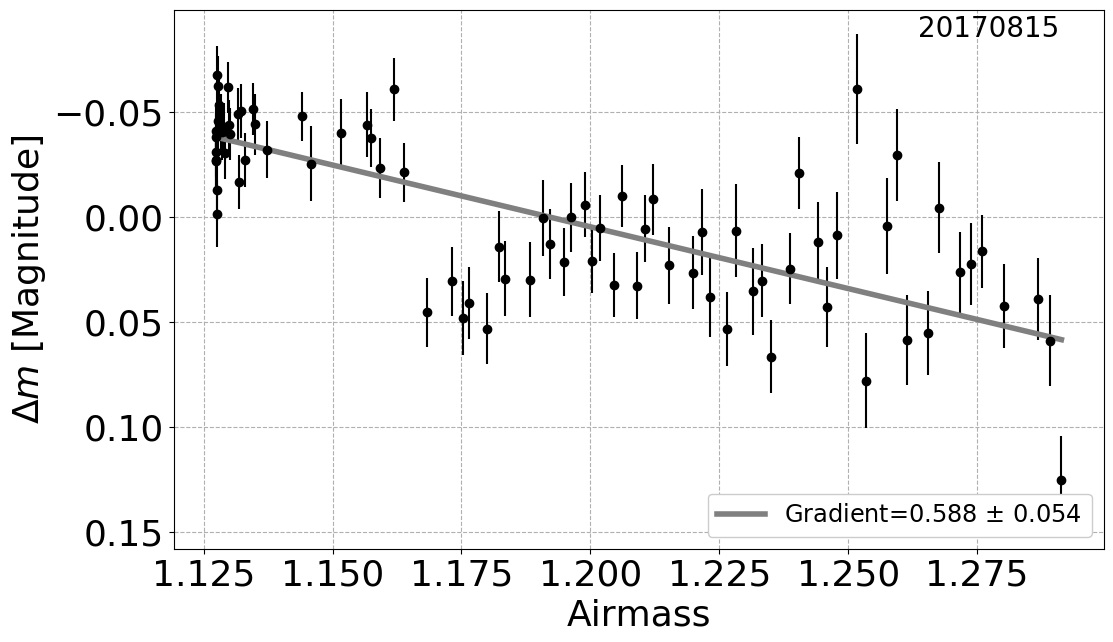}
    }
    \vskip\baselineskip
    \subfloat{
        \centering
        \includegraphics[width=0.48\textwidth, height=6cm]{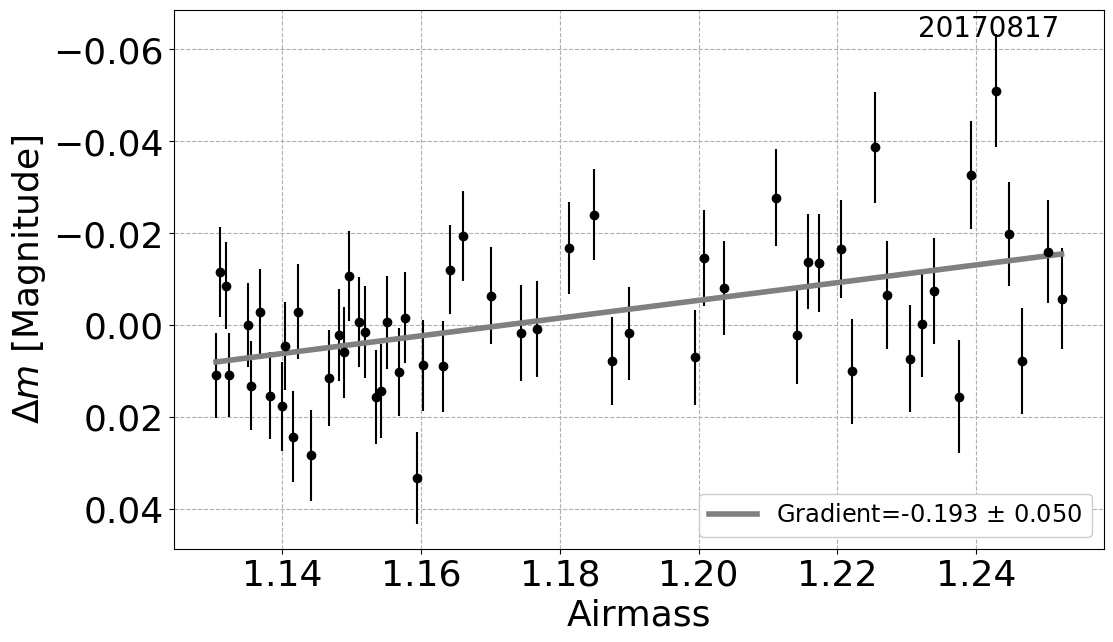}
    }
    \hfill
    \subfloat{  
        \centering 
        \includegraphics[width=0.48\textwidth, height=6cm]{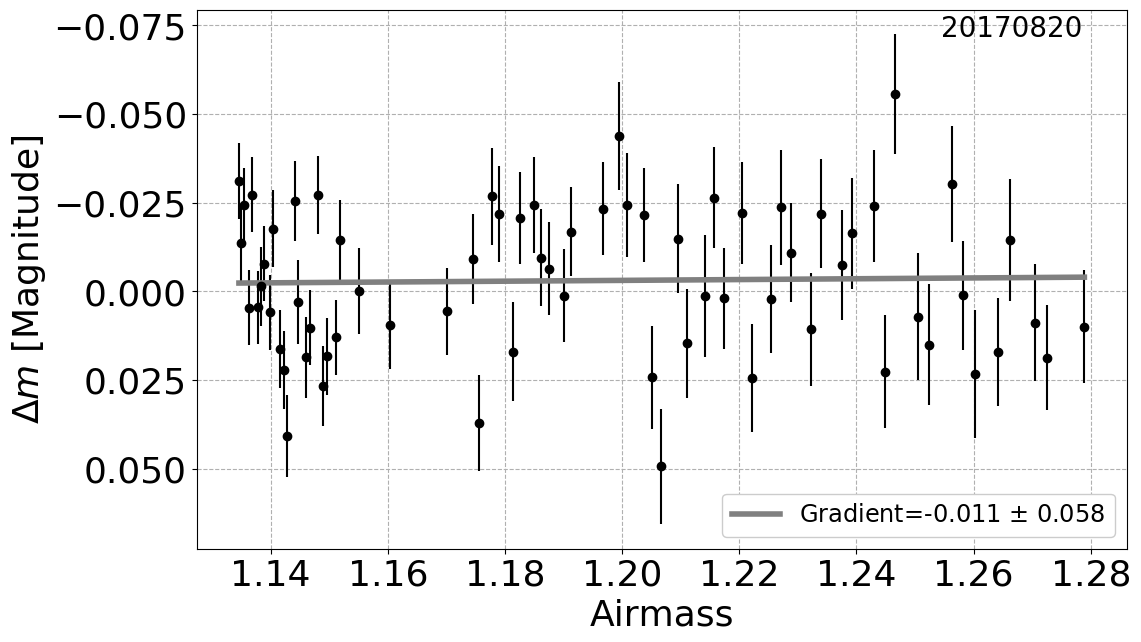}
    }
    \vskip\baselineskip
    \subfloat{
        \centering 
        \includegraphics[width=0.48\textwidth, height=6cm]{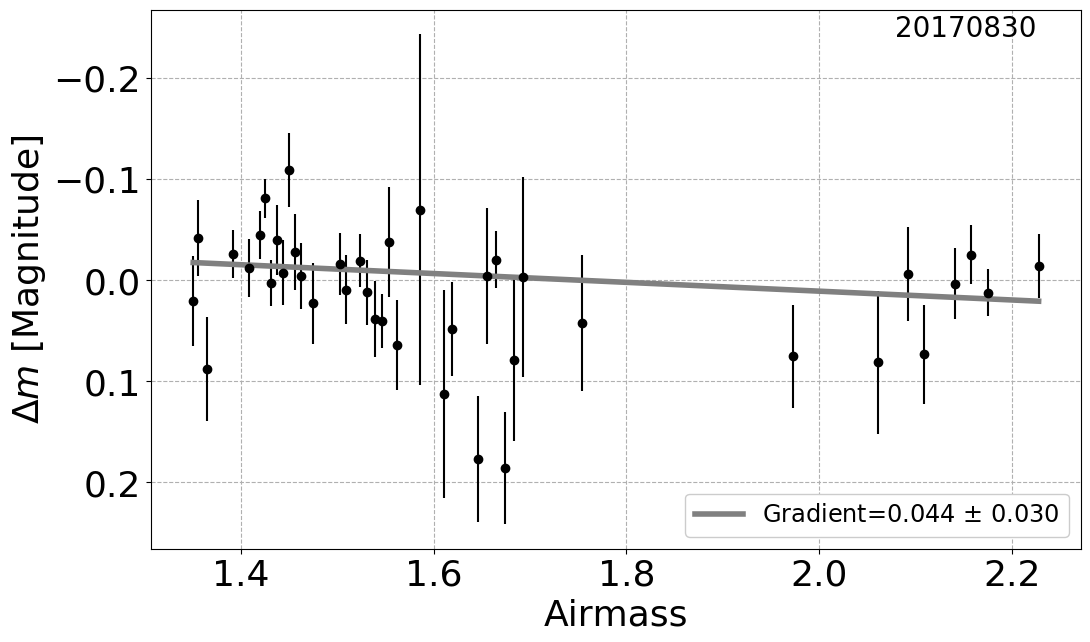}
    }
    \quad
    \subfloat{   
        \centering 
        \includegraphics[width=0.48\textwidth, height=6cm]{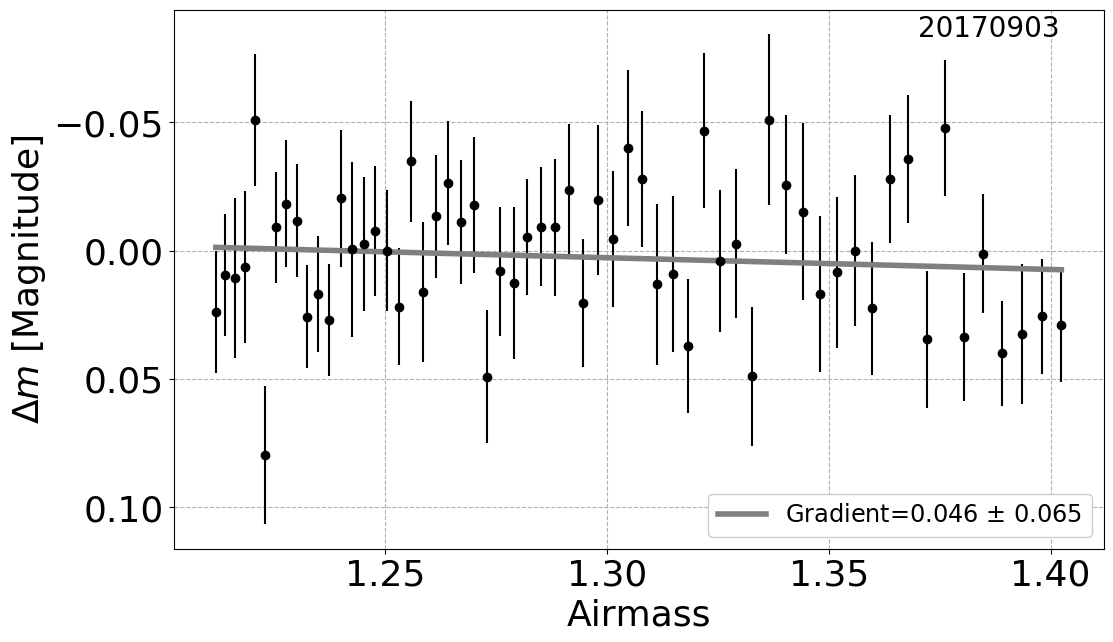}
    }
    \caption{Normalised differential photometry vs airmass for 6 additional nights not shown in Section \ref{sec:withinnight} (2017-07-24 to 2017-09-03). The gradient and corresponding bootstrap uncertainty of the plotted best fit straight lines, and the Pearson Correlation coefficients, are shown in the legends.}
\end{figure*}

\begin{figure*}
    \centering
    \subfloat{
        \centering
        \includegraphics[width=0.48\textwidth, height=6cm]{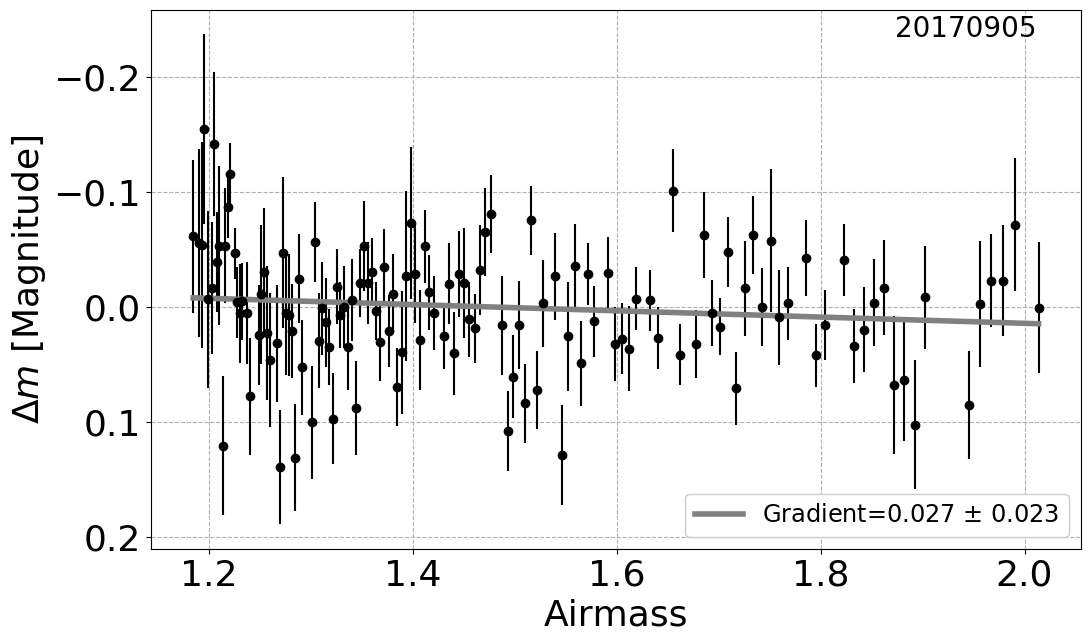}
    }
    \hfill
    \subfloat{
        \centering 
        \includegraphics[width=0.48\textwidth, height=6cm]{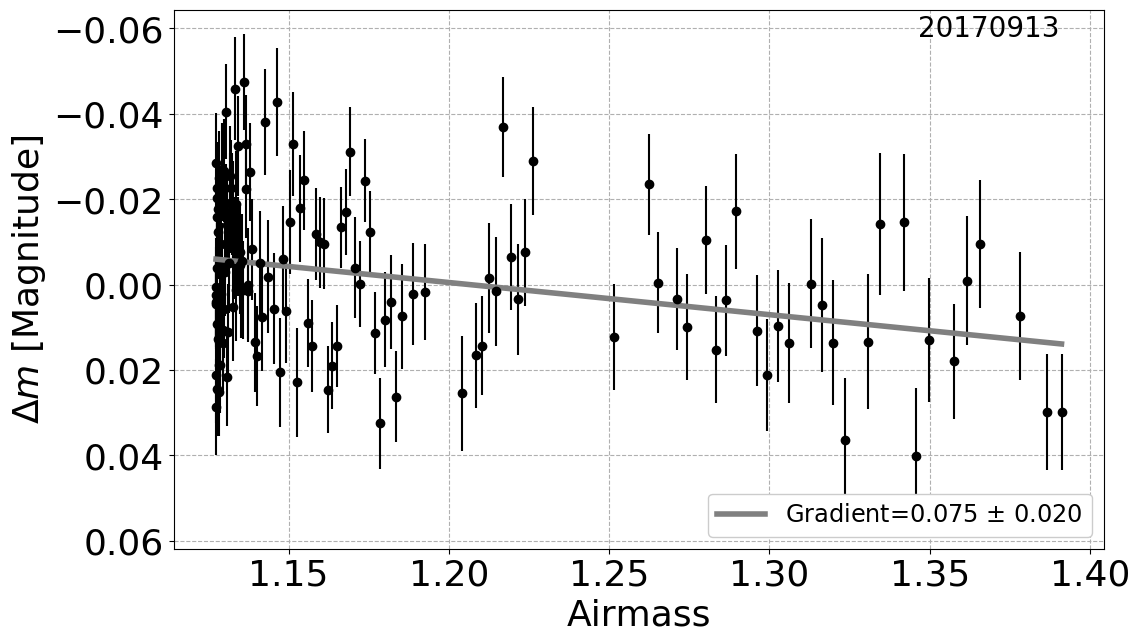}
    }
    \vskip\baselineskip
    \subfloat{
        \centering
        \includegraphics[width=0.48\textwidth, height=6cm]{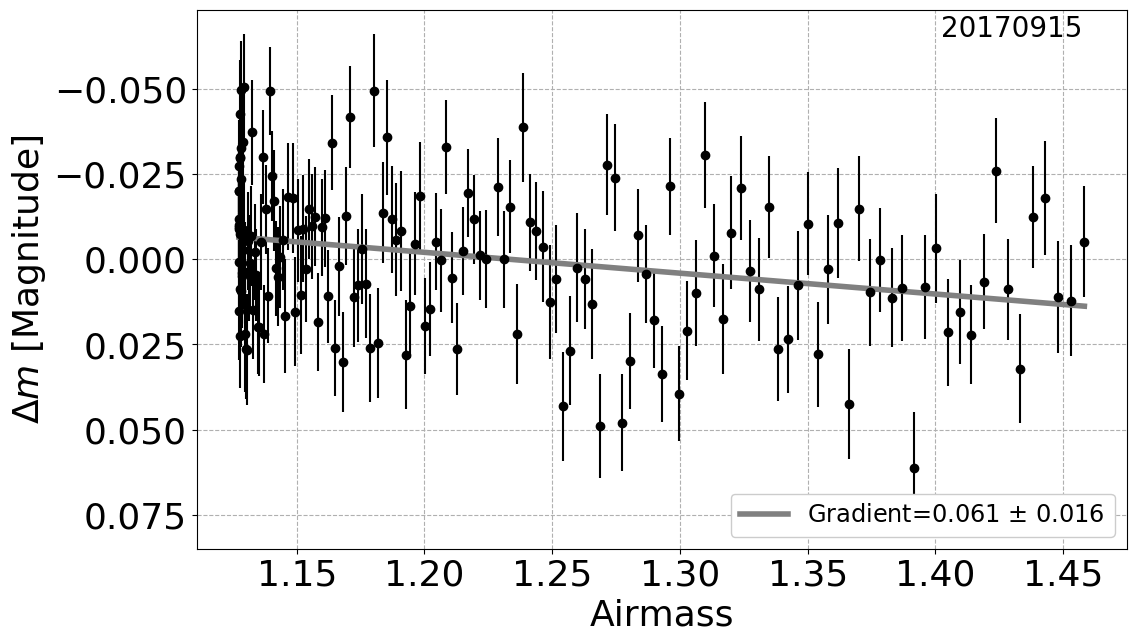}
    }
    \hfill
    \subfloat{  
        \centering 
        \includegraphics[width=0.48\textwidth, height=6cm]{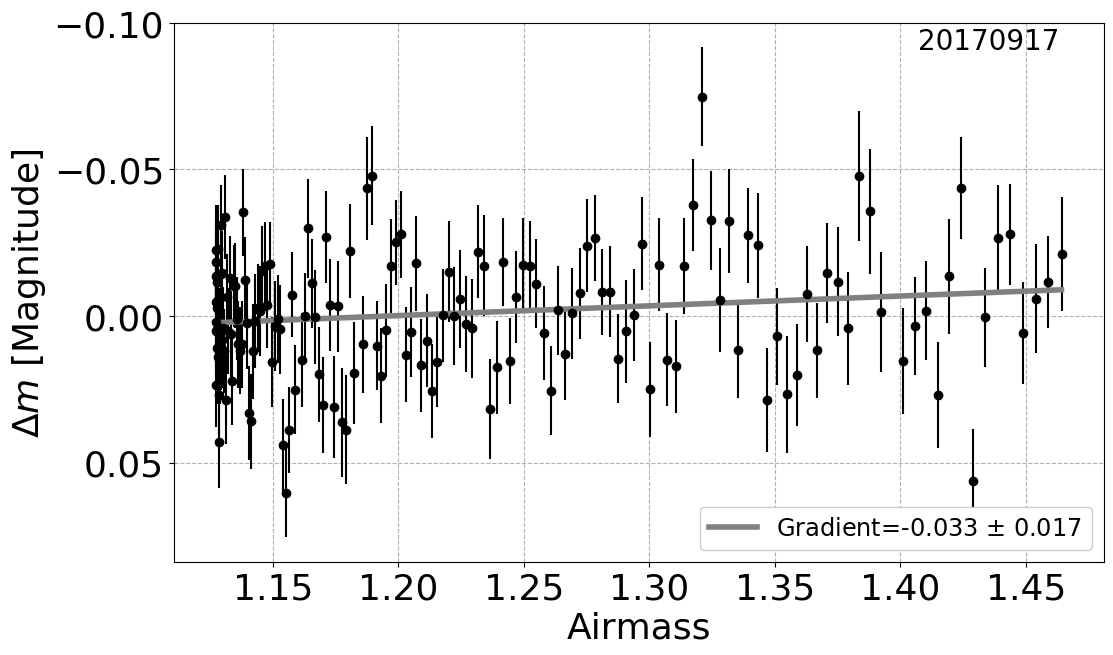}
    }
    \vskip\baselineskip
    \subfloat{
        \centering 
        \includegraphics[width=0.48\textwidth, height=6cm]{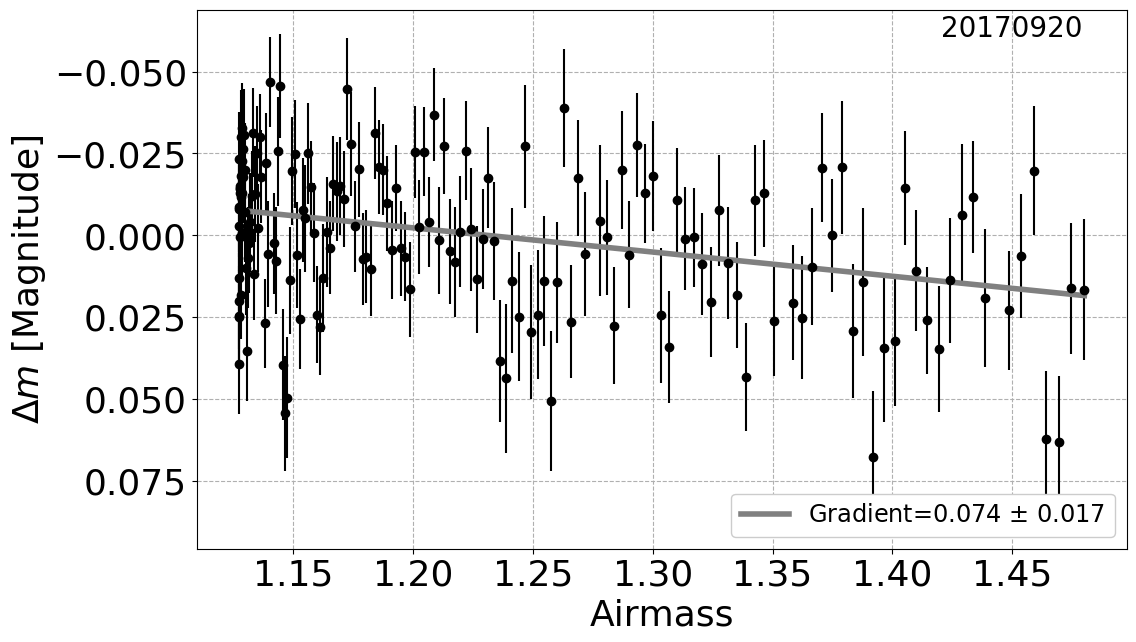}
    }
    \quad
    \subfloat{   
        \centering 
        \includegraphics[width=0.48\textwidth, height=6cm]{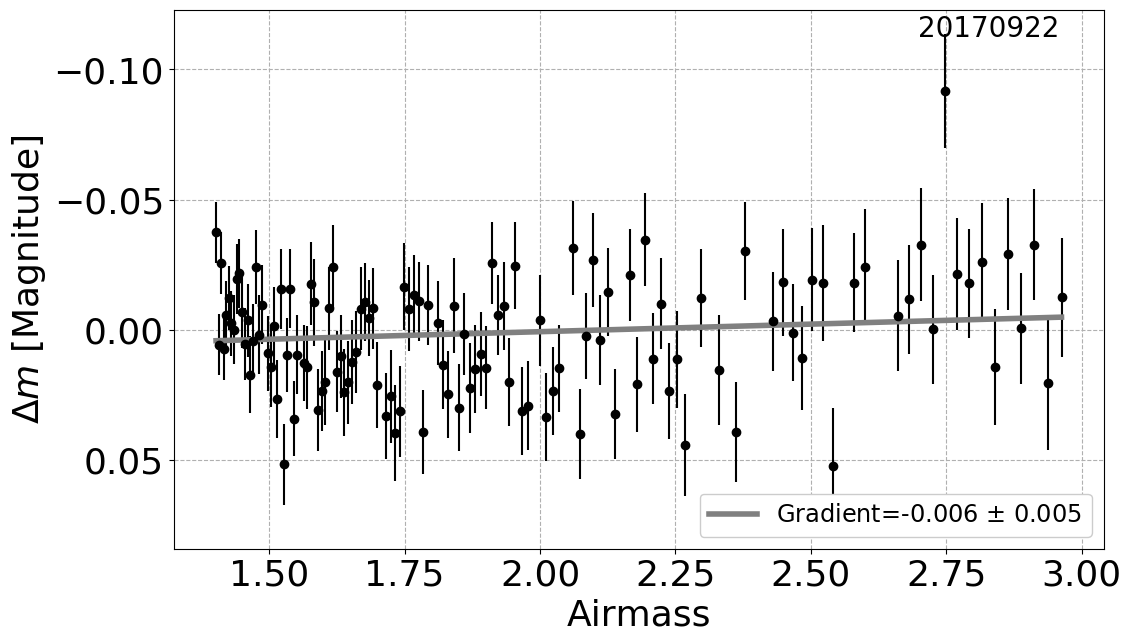}
    }
    \caption{Normalised differential photometry vs airmass for 6 additional nights not shown in Section \ref{sec:withinnight} (2017-09-05 to 2017-09-22). The gradient and corresponding bootstrap uncertainty of the plotted best fit straight lines, and the Pearson Correlation coefficients, are shown in the legends.}
\end{figure*}

\begin{figure*}
    \centering
    \subfloat{
        \centering
        \includegraphics[width=0.48\textwidth, height=6cm]{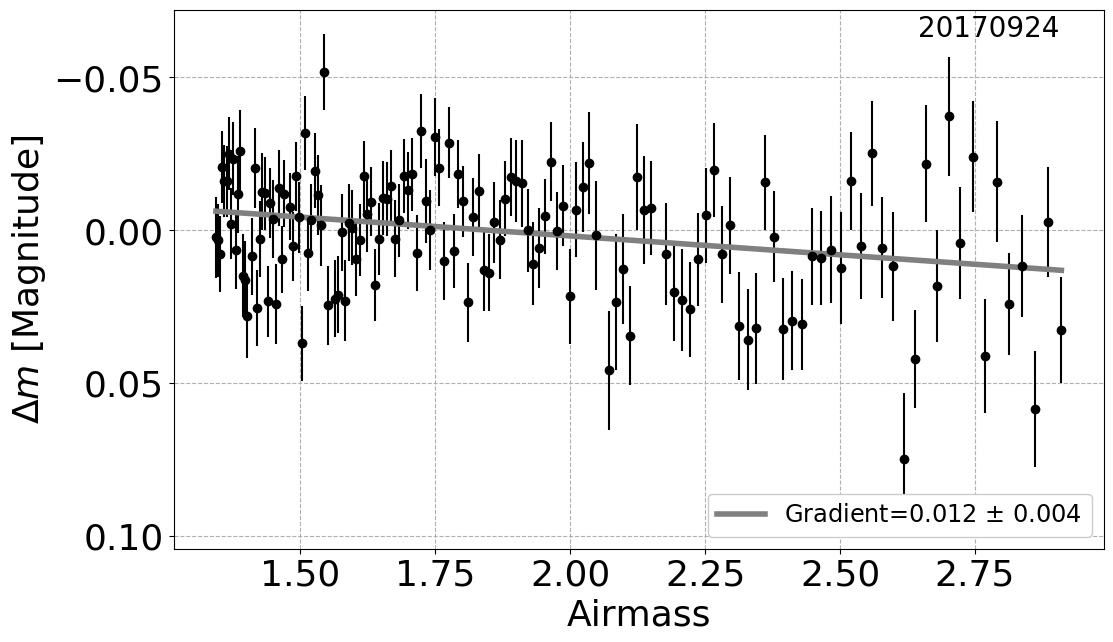}
    }
    \hfill
    \subfloat{
        \centering 
        \includegraphics[width=0.48\textwidth, height=6cm]{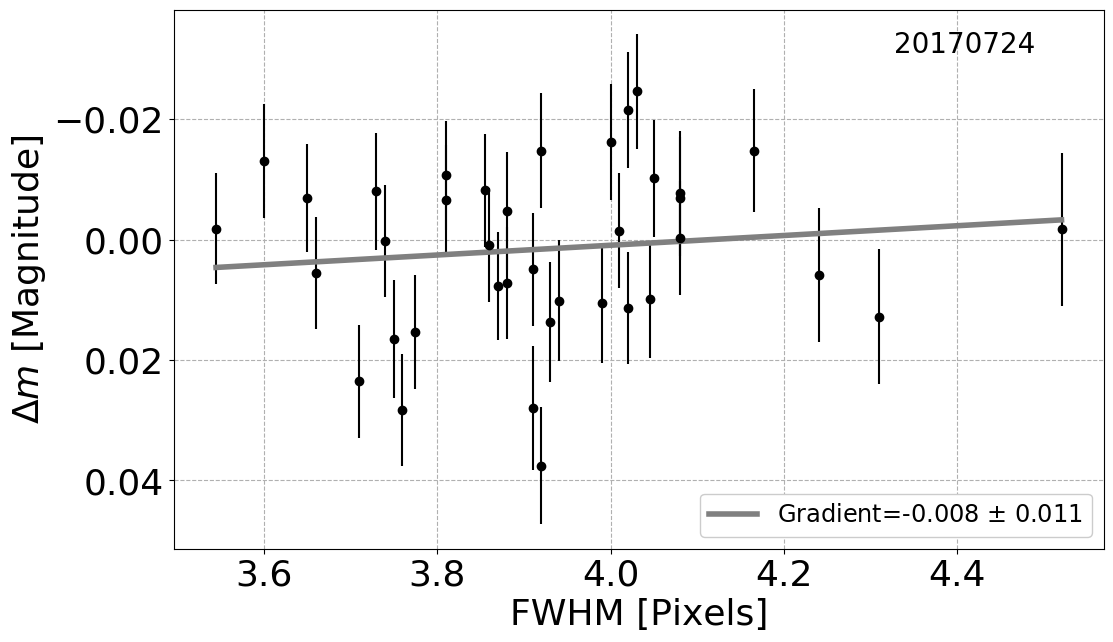}
    }
    \vskip\baselineskip
    \subfloat{
        \centering
        \includegraphics[width=0.48\textwidth, height=6cm]{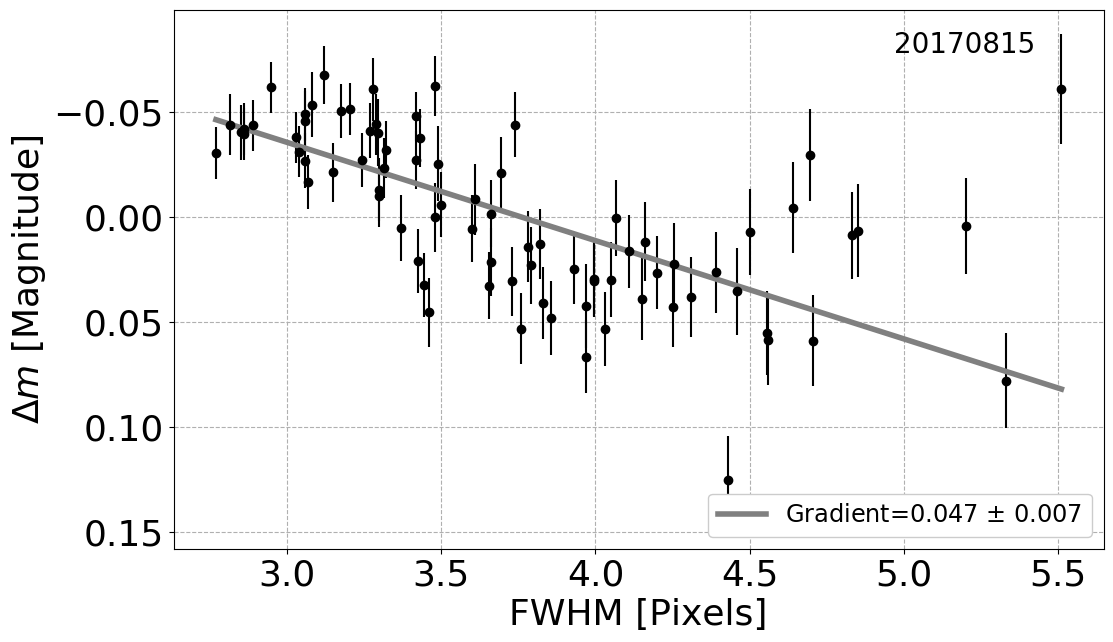}
    }
    \hfill
    \subfloat{  
        \centering 
        \includegraphics[width=0.48\textwidth, height=6cm]{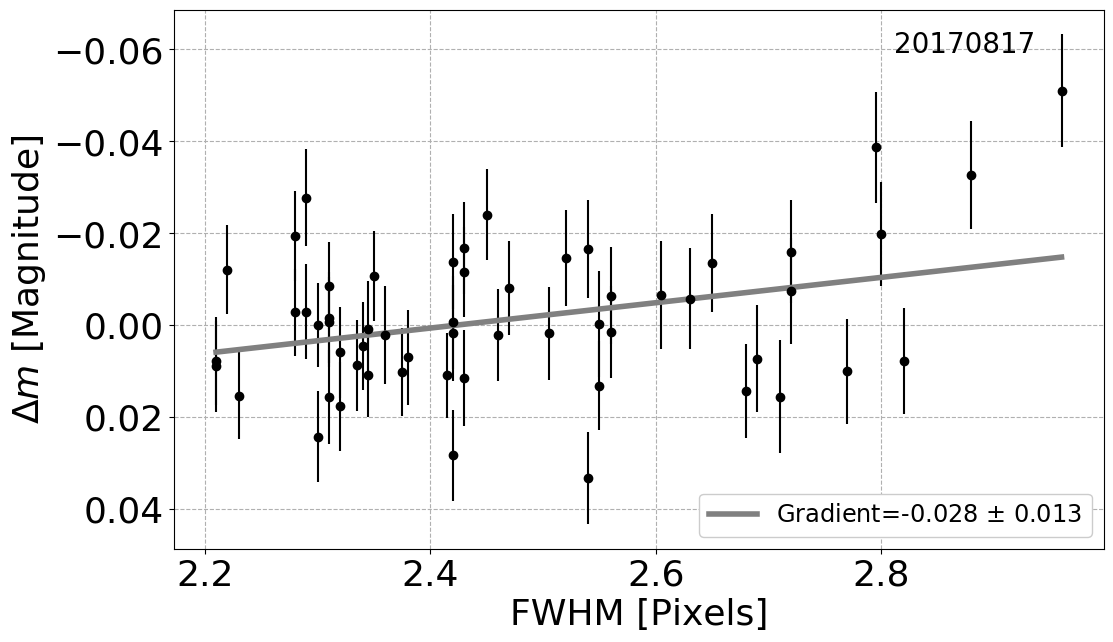}
    }
    \vskip\baselineskip
    \subfloat{
        \centering 
        \includegraphics[width=0.48\textwidth, height=6cm]{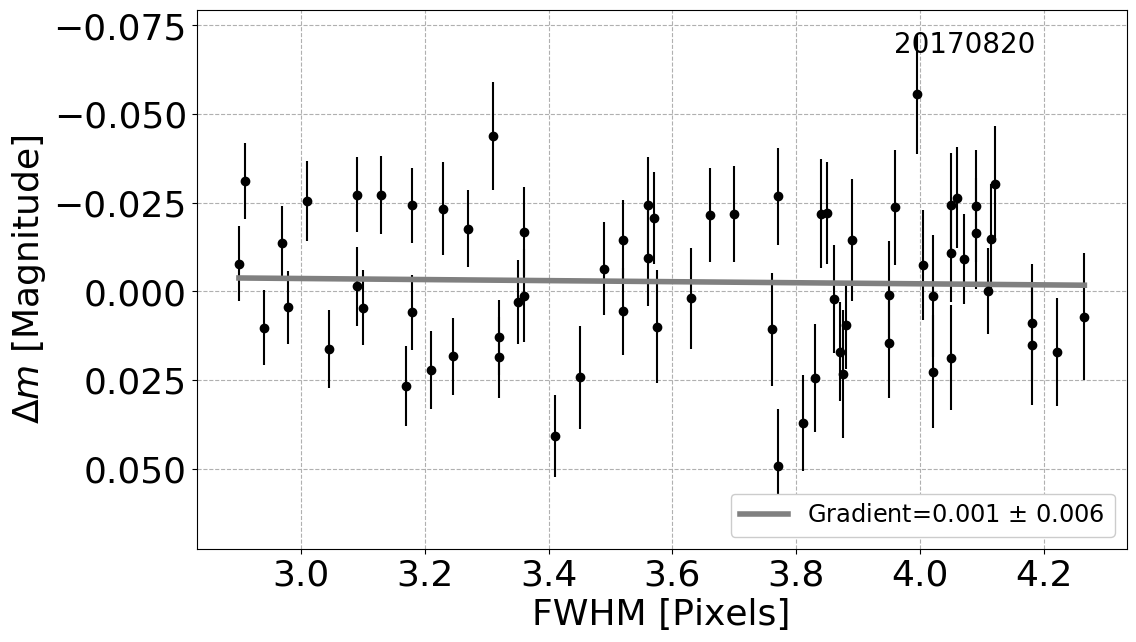}
    }
    \quad
    \subfloat{   
        \centering 
        \includegraphics[width=0.48\textwidth, height=6cm]{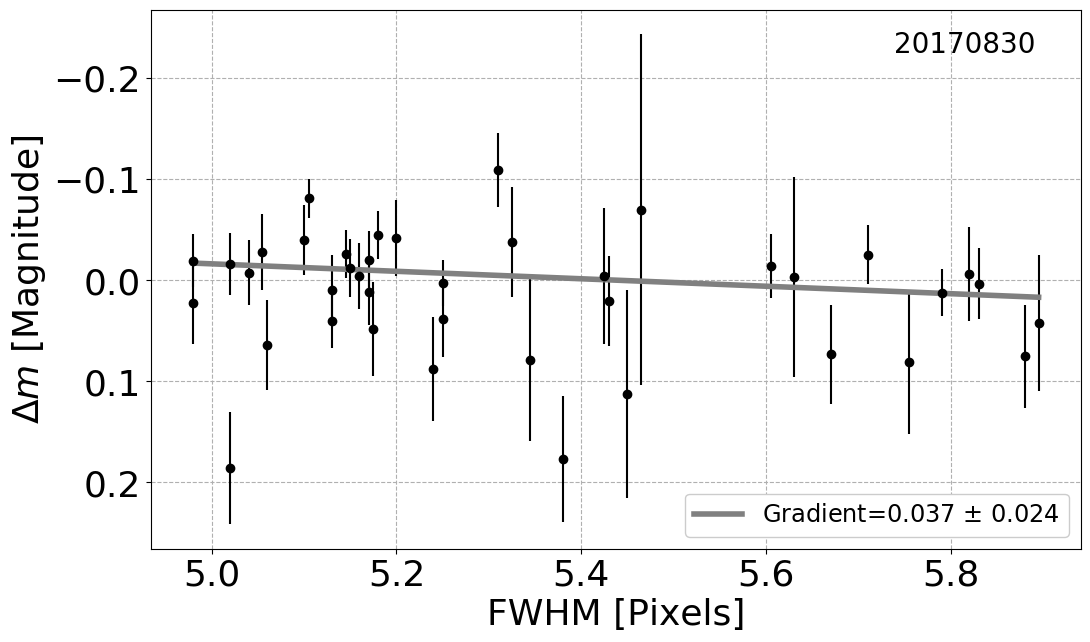}
    }
    \caption{(Top left) Normalised differential photometry vs airmass for the night of 2017-09-24. (Rest) Normalised differential photometry vs image FWHM (as a proxy for seeing) for 5 additional nights not shown in Section \ref{sec:withinnight} (2017-07-24 to 2017-08-30). The gradient and corresponding bootstrap uncertainty of the plotted best fit straight lines, and the Pearson Correlation coefficients, are shown in the legends.}
\end{figure*}

\begin{figure*}
    \centering
    \subfloat{
        \centering
        \includegraphics[width=0.48\textwidth, height=6cm]{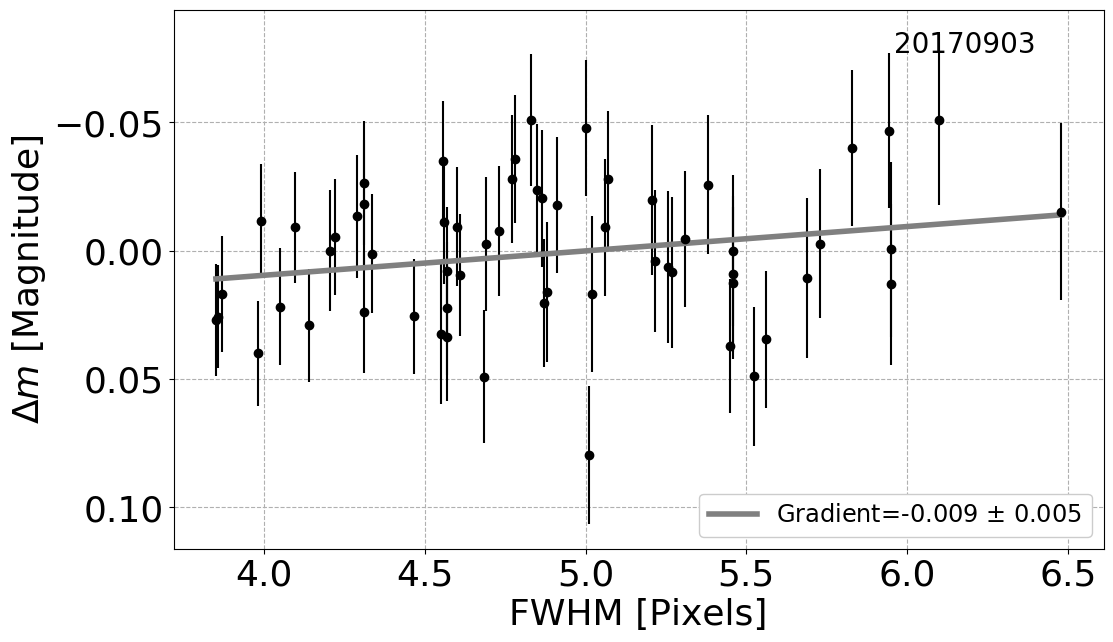}
    }
    \hfill
    \subfloat{
        \centering 
        \includegraphics[width=0.48\textwidth, height=6cm]{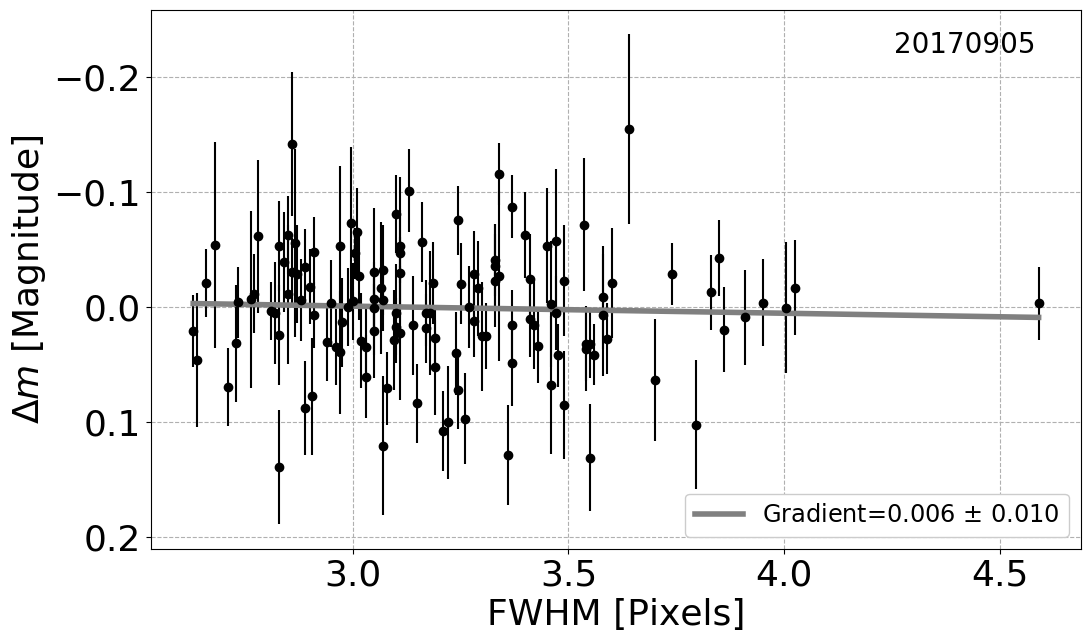}
    }
    \vskip\baselineskip
    \subfloat{
        \centering
        \includegraphics[width=0.48\textwidth, height=6cm]{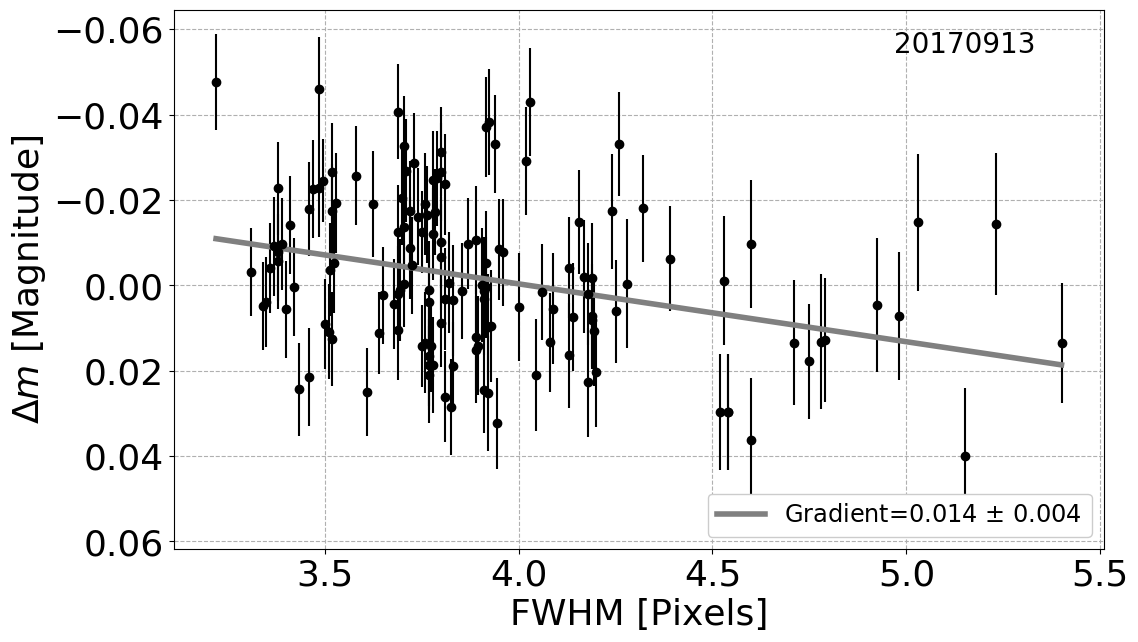}
    }
    \hfill
    \subfloat{  
        \centering 
        \includegraphics[width=0.48\textwidth, height=6cm]{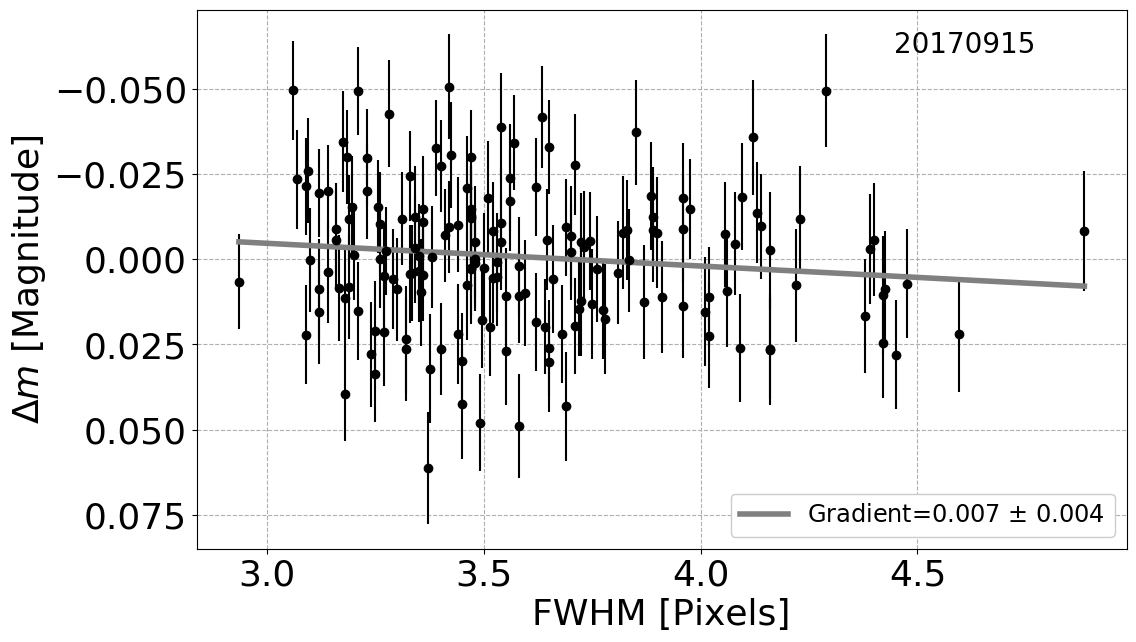}
    }
    \vskip\baselineskip
    \subfloat{
        \centering 
        \includegraphics[width=0.48\textwidth, height=6cm]{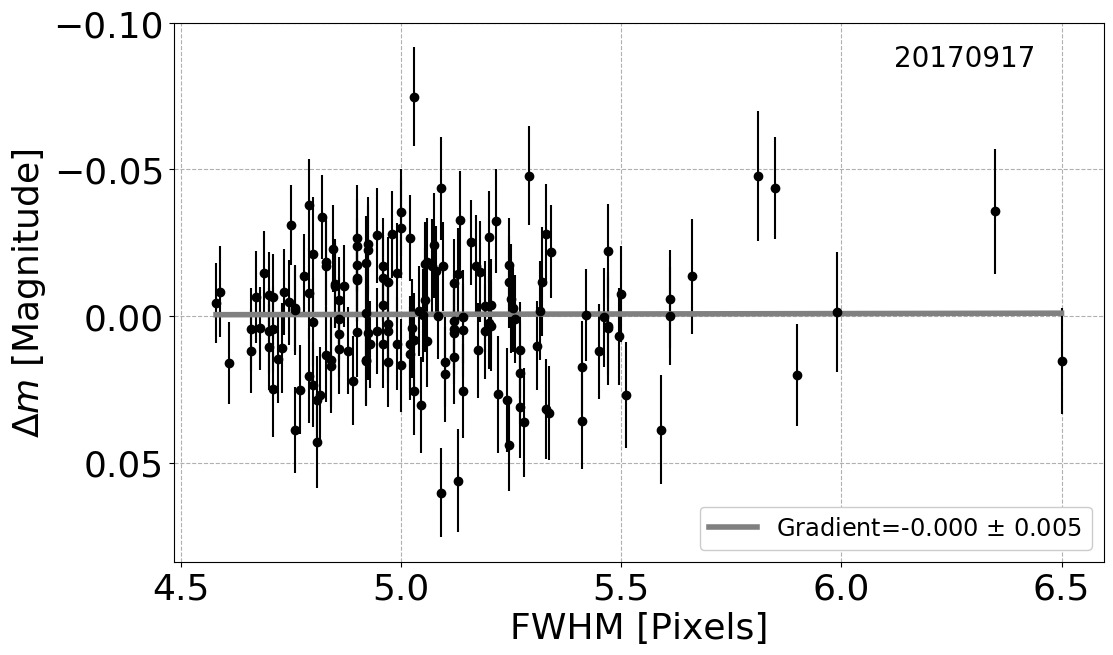}
    }
    \quad
    \subfloat{   
        \centering 
        \includegraphics[width=0.48\textwidth, height=6cm]{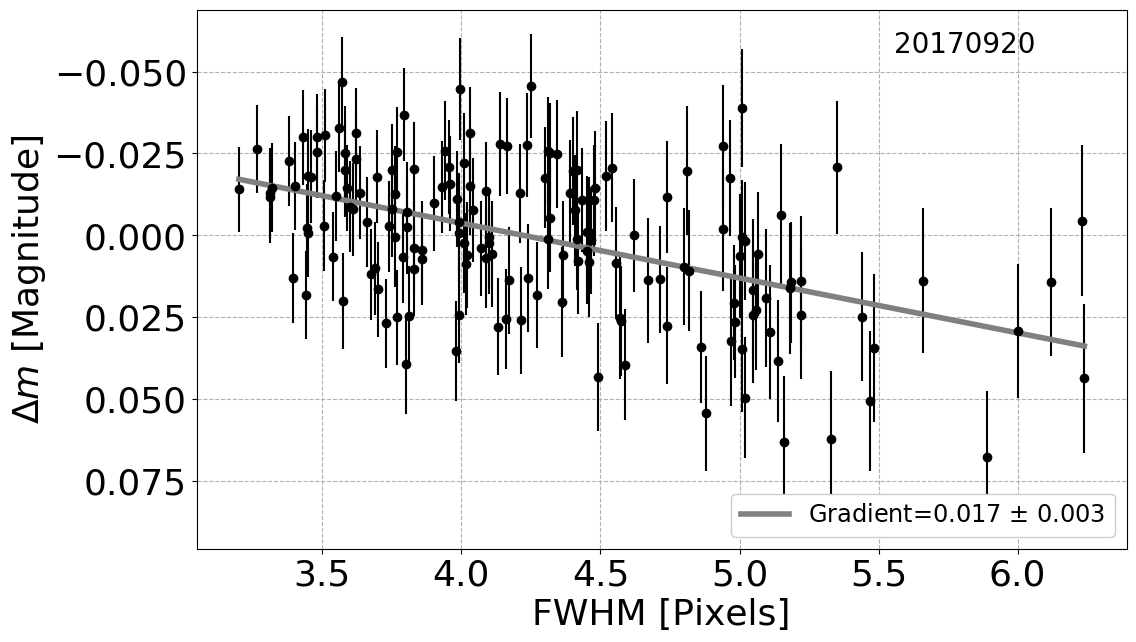}
    }
    \caption{Normalised differential photometry vs image FWHM (as a proxy for seeing) for 6 additional nights not shown in Section \ref{sec:withinnight} (2017-09-03 to 2017-09-20). The gradient and corresponding bootstrap uncertainty of the plotted best fit straight lines, and the Pearson Correlation coefficients, are shown in the legends.}
\end{figure*}

\begin{figure*}
    \centering
    \subfloat{
        \centering 
        \includegraphics[width=0.48\textwidth, height=6cm]{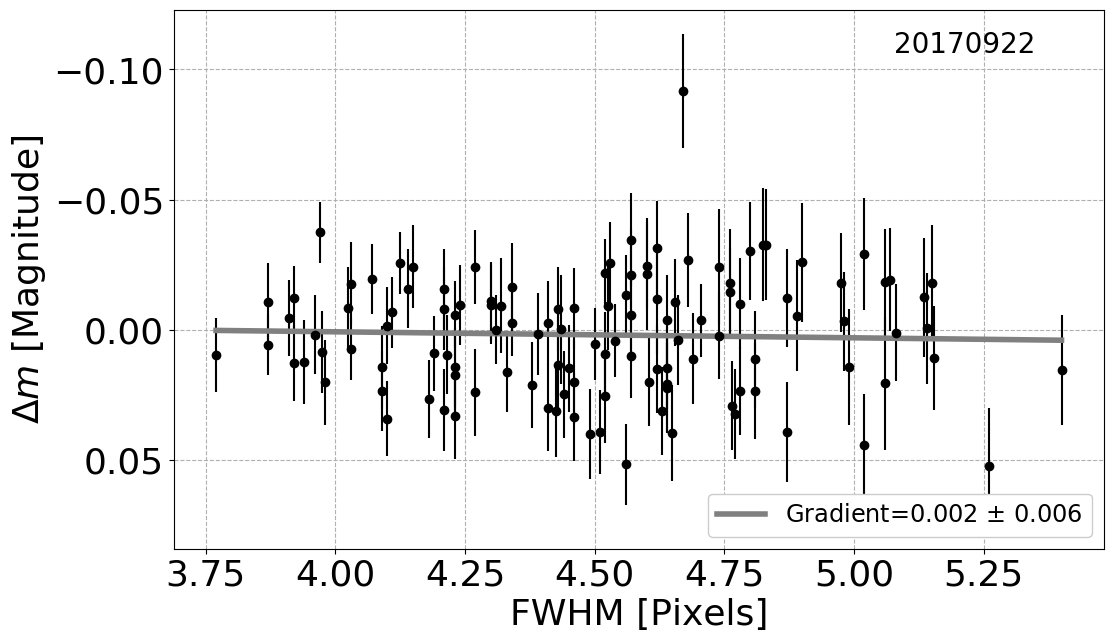}
    }
    \quad
    \subfloat{   
        \centering 
        \includegraphics[width=0.48\textwidth, height=6cm]{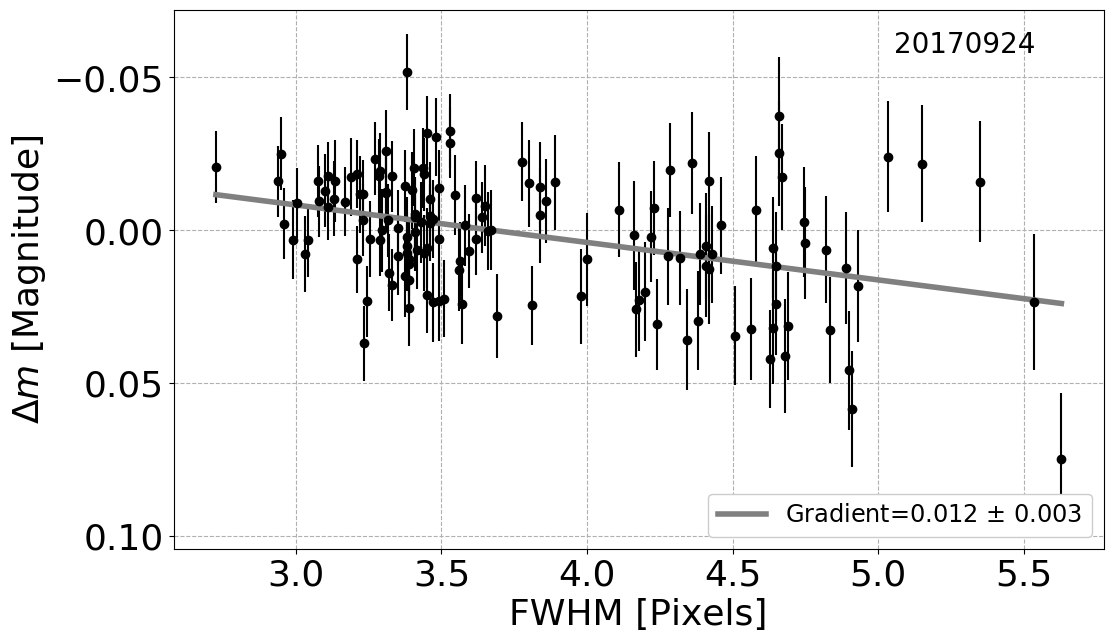}
    }
    \caption{Normalised differential photometry vs image FWHM (as a proxy for seeing) for 2 additional nights not shown in Section \ref{sec:withinnight} (2017-09-22 to 2017-09-24). The gradient and corresponding bootstrap uncertainty of the plotted best fit straight lines, and the Pearson Correlation coefficients, are shown in the legends.}
\end{figure*}

\section{The observable signal strength of lightning flashes}\label{sec:lightningeqs}

\textcolor{black}{In Equation \ref{eq:Iobs} and \ref{eq:Iopt} below, all fluxes, $I$, are in units of Jansky (Jy) -- where 1 W.s.m$^-2$ = $10^{26}$ Jy -- and all other variables are in SI.}

\textcolor{black}{Following \citet{hodosan2017lightning}, the total observable flux from a lightning storm on the surface of a brown dwarf may be expressed as}

\begin{equation}
    I_{\textrm{obs}} = I_{\textrm{opt,fl}}\frac{\tau_{\textrm{fl}}}{\tau_{\textrm{obs}}}n_{\textrm{tot, fl}}\;,
    \label{eq:Iobs}
\end{equation}

\textcolor{black}{where $I_{\textrm{opt,fl}}$ is the optical flux from a single strike, $\tau_{\textrm{fl}}$ is the duration of the strike, $\tau_{\textrm{obs}}$ is the exposure time and $n_{\textrm{tot, fl}}$ is the total number of observed flashes. Next, we can write $I_{\textrm{opt,fl}}$ as}

\begin{equation}
    I_{\textrm{opt,fl}} = \frac{P_{\textrm{opt,fl}}}{f_{\textrm{eff}}}\frac{10^{26}}{4\pi d^2}\;,
    \label{eq:Iopt}
\end{equation}

\textcolor{black}{when observing in a filter with effective frequency $f_\textrm{eff}$, for a flash with an optical power of $P_{\textrm{opt,fl}}$ at a distance of $d$. Finally, the total number of flashes over the visible hemisphere of the brown dwarf can be written as}

\begin{equation}
    n_{\textrm{tot, fl}} = \rho_{\textrm{fl}}\times2\pi R^2 \times\tau_\textrm{obs}\;
    \label{eq:ntot}
\end{equation}

for a flash density of $\rho_{\textrm{fl}}$, and a radius of $R$, which is assumed to be $1R_{\textrm{Jupiter}}$. Subsequently, any estimated fluxes can be converted to magnitudes via Pogson's Formula,
\begin{equation}
    m - m_{\textrm{zp}} = -2.5\log\frac{F}{F_{\textrm{zp}}}\;,
    \label{eq:pogson}
\end{equation}

\textcolor{black}{where the relevant magnitude and flux zero-points (i.e. $m_{\textrm{zp}}$ and $F_{\textrm{zp}}$) for the Johnson-Cousin filters used in this work are taken from \citet{bessell1990ubvri} and \citet{bessell1998model} respectively. We again refer the reader to \citet{hodosan2017lightning} for full details of the parameters -- both instrumental and physical -- used in this work.}

\textcolor{black}{Substituting Equation \ref{eq:ntot} into Equation \ref{eq:Iobs} shows that the exposure times cancel, and so $I_{\textrm{obs}}$ only depends on the duration of the lightning discharge. This at first seems strange, but note that here we are not trying to detect a \textit{single} strike during an otherwise flat time series -- where a short exposure time really would be of benefit -- but rather this is simply a measure of any net brightening of the source due to the presence of a lightning storm i.e. very many strikes occurring collectively, with a rate per unit area described by the flash density, during any given exposure.}

\section{Affiliations}\label{sec:affiliations}

$^{1}$ Centre for Exoplanet Science, University of St Andrews, North Haugh, KY16 9SS, UK\\
$^{2}$ SUPA, School of Physics and Astronomy, University of St. Andrews, North Haugh, KY16 9SS, UK\\
$^{3}$ SRON Netherlands Institute for Space Research, Sorbonnelaan 2, 3584 CA Utrecht, NL \\
$^{4}$Niels Bohr Institute \& Centre for Star and Planet Formation, University of Copenhagen, {\O}ster Voldgade 5, 1350 Copenhagen, Denmark   \\
$^{5}$Dipartimento di Fisica "E.R. Caianiello", Universit{\`a} di Salerno, Via Giovanni Paolo II 132, 84084, Fisciano, Italy   \\
$^{6}$Istituto Nazionale di Fisica Nucleare, Sezione di Napoli, Napoli, Italy   \\
$^{7}$Astronomisches Rechen-Institut, Zentrum f{\"u}r Astronomie der Universit{\"a}t Heidelberg (ZAH), 69120 Heidelberg, Germany   \\
$^{8}$Centre for Astrophysics \&\ Planetary Science, The University of Kent, Canterbury CT2 7NH, UK\\
$^{9}$Department~of~Physics,~Isfahan~University~of~Technology,~Isfahan~84156-83111,~Iran   \\
$^{10}$Universit{\"a}t Hamburg, Faculty of Mathematics, Informatics and Natural Sciences, Department of Earth Sciences, \\ Meteorological Institute, Bundesstra\ss{}e 55, 20146 Hamburg, Germany   \\ 
$^{11}$Astrophysics Group, Keele University, Staffordshire, ST5 5BG, UK   \\
$^{12}$Institute for Advanced Research, Nagoya University, Furo-cho, Chikusa-ku, Nagoya, 464-8601, Japan   \\
$^{13}$Instituto de Astronomia y Ciencias Planetarias de Atacama,  Universidad de Atacama, Copayapu 485,  Copiapo, Chile \\
$^{14}$Max Planck Institute for Astronomy, K{\"o}nigstuhl 17, 69117 Heidelberg, Germany \\
$^{15}$Chungnam National University, Department of Astronomy and Space Science, 34134 Daejeon, Republic of Korea   \\
$^{16}$Department of Physics, University of Rome ``Tor Vergata'', Via della Ricerca Scientifica 1, I-00133, Rome, Italy   \\
$^{17}$Unidad de Astronom{\'{\i}}a, Universidad de Antofagasta, Av.\ Angamos 601, Antofagasta, Chile  \\ 
$^{18}$Department of Physics, Sharif University of Technology, PO Box 11155-9161 Tehran, Iran   \\
$^{19}$Las Cumbres Observatory Global Telescope, 6740 Cortona Dr., Suite 102, Goleta, CA 93111, USA\\
$^{20}$Department of Physics, University of California, Santa Barbara, CA 93106-9530, USA\\
$^{21}$Centre for Electronic Imaging, Department of Physical Sciences, The Open University, Milton Keynes, MK7 6AA, UK   \\
$^{22}$Institute for Astronomy, University of Edinburgh, Royal Observatory, Edinburgh EH9 3HJ, UK \\
$^{23}$Stellar Astrophysics Centre, Department of Physics and Astronomy, Aarhus University, Ny Munkegade 120, 8000 Aarhus C, Denmark \\
$^{24}$\,INAF -- Osservatorio Astrofisico di Torino, via Osservatorio
20, I-10025, Pino Torinese, Italy \\
$^{25}$\,International Institute for Advanced Scientific Studies
(IIASS), Via G. Pellegrino 19, I-84019, Vietri sul Mare (SA), Italy \\
$^{26}$\,SUPA, School of Science and Engineering, University of Dundee, Nethergate, Dundee DD1 4HN, U.K \\
$^{27}$\,Department of Physics, Nagoya University, Furo-cho, Chikusa-ku, Nagoya, 464-8602, Japan \\
$^{28}$\,Niels Bohr International Academy, The Niels Bohr Institute, Blegdamsvej 17, DK-2100, Copenhagen, Denmark \\
$^{29}$\,Facultad de Ingenier{\'{\i}}a y Tecnolog{\'{\i}}a, Universidad San Sebastian, General Lagos 1163, Valdivia 5110693, Chile \\
$^{30}$\,Facultad de Econom{\'{\i}}a y Negocios, Universidad San Sebastian, General Lagos 1163, Valdivia 5110693, Chile


\bsp	
\label{lastpage}
\end{document}